\documentclass[twocolumn,twocolappendix,tighten]{aastex6}

\usepackage{color}
\usepackage{amsmath}
\usepackage{natbib}
\slugcomment{ApJ manuscript, revised version \today}

\shorttitle{Testing the cosmological redshift of galaxy spectra}
\shortauthors{Ferreras \& Trujillo}

\begin{document}

\title{Testing the wavelength dependence of cosmological redshift  down to $\Delta$z$\sim 10^{-6}$}


\author{Ignacio Ferreras\altaffilmark{1,2}}
\affil{Mullard Space Science Laboratory, University College
  London, Holmbury St Mary, Dorking, Surrey RH5 6NT, UK}
\author{Ignacio Trujillo}
\affil{Instituto de Astrof\'isica de Canarias, E-38200 La Laguna,
  Tenerife, Spain\\
  Departamento de Astrof\'isica, Universidad de La Laguna, E-38205
  La Laguna, Tenerife, Spain}

\altaffiltext{1}{i.ferreras@ucl.ac.uk}
\altaffiltext{2}{Severo Ochoa visitor. Instituto de Astrof\'\i sica de Canarias.}

\vskip1truecm
\centerline{Accepted for publication in {\it The Astrophysical Journal}, May 2016}

\begin{abstract}
At the core of the standard cosmological model lies the assumption
that the redshift of distant galaxies is independent of photon
wavelength. This invariance of cosmological redshift with wavelength
is routinely found in all galaxy spectra with a precision of $\Delta
z\sim 10^{-4}$. The combined use of approximately half a million
high-quality galaxy spectra from the Sloan Digital Sky Survey (SDSS)
allows us to explore this invariance down to a nominal precision in
redshift of $10^{-6}$ (statistical).  Our analysis is performed over
the redshift interval 0.02$<$z$<$0.25. We use the centroids of
spectral lines over the 3700--6800\AA\ rest-frame optical window.  We
do not find any difference in redshift between the blue and red sides
down to a precision of $10^{-6}$ at z$\lesssim$0.1 and $10^{-5}$ at
0.1$\lesssim$z$\lesssim$0.25 (i.e. at least an order of magnitude
better than with single galaxy spectra). This is the first time the
wavelength-independence of the (1+z) redshift law is confirmed over a
wide spectral window at this precision level.  This result holds
independently of the stellar population of the galaxies and their
kinematical properties. This result is also robust against wavelength
calibration issues. The limited spectral resolution (R$\sim$2000) of
the SDSS data combined with the asymmetric wavelength sampling of the
spectral features in the observed restframe due to the (1+z)
stretching of the lines prevent our methodology to achieve a precision
higher than $10^{-5}$, at z$>$0.1. Future attempts to
constrain this law will require high quality galaxy spectra at higher
resolution ($R\gtrsim 10,000$).
\end{abstract}


\section{Introduction}
\label{sec:intro}

Since the seminal work of \citet{HH:31} on the recession
velocities of extragalactic nebul\ae , the connection between the
redshift of distant sources and the evolution of the
Universe has been the central pillar of observational
cosmology. Due to the expanding nature of our Universe, light from
distant sources is redshifted following a simple scaling law
$\lambda_{\rm obs}=\lambda_0(1+z)$, where $\lambda_{\rm obs}$ is the
observed wavelength and $\lambda_0$ is the wavelength in the rest
frame of the source. Although challenged in the past as a measurement
of the expansion of the Universe \citep{Arp:87}, the discovery of such a trend
over millions of galaxy spectra  out to z$\gtrsim 7$ \citep[e.g. ][]{Oesch:15} has been one
of the major successes of our cosmological framework.

According to the (1+z) wavelength scaling law, every spectral line
from an astronomical object is redshifted by the same factor. Current
individual optical galaxy spectra allow us to explore this stretching
effect with an accuracy of the order of $10^{-4}$. So far, no evidence has
been found regarding a deviation from the theoretical prediction at
this level of precision. However, with the advent of large
spectroscopic galaxy surveys, such as the Sloan Digital Sky Survey
\citep[SDSS,][]{SDSS}, containing millions of spectra, it is possible
to explore the cosmological redshift law, statistically achieving a
much higher accuracy. In fact, by combining the information from all
available, high-quality galaxy spectra in SDSS, it is possible to
explore any departure from the predicted, wavelength-independent, trend
by a factor 100 times more accurately than with individual galaxy
spectra.

The aim of this paper is to quantify any deviation from the standard
cosmological redshift law in galaxy spectra when the analysis is
conducted with a (statistical) precision of 1\,part-per-million.
Moreover, we assess whether such deviation shows any trend with
redshift or wavelength.  If this were the case, a variation of the
(1+z) law with cosmic time could be suggestive of new (unexplored)
physical, or astrophysical phenomena. A number of potential mechanisms
can be found in the literature that would produce a dependence of the
redshift of the galaxies as a function of the observed wavelength. For
instance, \citet{Laio:97} predicted a quantum global effect of the
intergalactic plasma with the photons, changing the energy of the
photons without altering their trajectory.  Hence, a cumulative effect
along the line of sight caused by the interaction of light with the
ionized material of the intergalactic medium is a potential
scenario. With regards to a possible cosmological origin, one could
consider changes in the fundamental laws governing null geodesics in a
cosmological background that would introduce a chromatic term,
breaking Lorentz invariance \citep[introducing a wavelength-dependent
  speed of light][]{Nick:98,Abdo:09}.  Alternatively, a change in the
laws that govern the atomic emission and absorption processes
\citep{Webb:01,Quast:04,Rahmani:14,Albareti:15} could be responsible
for a wavelength-dependent redshift. Finally, it is worth mentioning
that well-known physical mechanisms could produce a non-cosmological
redshift, such as the Wolf effect \citep{Wolf:87}, or gravitational
redshift. The latter, at the level of accuracy presented here, could
be significant. For instance, the expected variation between the
gravitational redshift of a photon emitted at the center of a massive
($10^{11}$M$_\odot$) galaxy, and a photon from a region located 1\,kpc
away from the center should be $\Delta z\sim 5\times 10^{-6}$.

To pursue these goals, we focus on a wide spectral range in the
optical rest-frame window (3700--6800\AA), and adopt the following
ansatz:

\begin{equation} 
\lambda_{\rm    obs}=\lambda_0(1+z)[1+\Delta(z,\lambda_{0})], 
\label{eq:Deltaz} 
\end{equation}

where the departure is quantified by a non-zero
$\Delta(z,\lambda_{0})$, for which both a redshift and wavelength
dependence is considered.  We make use of the excellent database
provided by the Sloan Digital Sky Survey \citep[SDSS,][]{SDSS}, from
which we select galaxy spectra with a high signal-to-noise ratio
(S/N$>$15), out to redshift z$\lesssim$0.25, within which a large
number of high quality spectroscopic data can be extracted from SDSS.
For each spectrum it is possible to derive the observed wavelengths of
several prominent absorption and emission features. Although individual SDSS
spectral lines can only give redshifts with an accuracy between $10$
and $30$\,km\,s$^{-1}$ \citep{SDSS:DR1}, by combining tens of
thousands of high-quality spectra, it is possible to achieve a much
higher precision ($\sim 0.1$\,km\,s$^{-1}$).  \\

This paper presents a test of the wavelength dependence of
cosmological redshift, at the highest accuracy to date, with a
statistical uncertainty of $\Delta$z$\sim 10^{-6}$.  In
\S\ref{sec:sample} we describe the spectroscopic dataset.
\S\ref{sec:method} presents the methodology adopted to derive
redshifts with a number of carefully selected features.  The analysis
can be subject to various systematic effects from the instrumentation,
the data reduction process, or the source selection. We explore the
effect of stellar populations on the derivation of the line centroids
in \S\ref{sec:agemet}. Additional systematic effects are discussed in
an Appendix. Finally \S\ref{sec:discussion} briefly summarizes the
results.

\section{Sample Selection}
\label{sec:sample}

Our analysis is based on the SDSS spectroscopic sample. We select from
Data Release 10 all individual galaxy spectra \citep{SDSS:DR10} with a
median signal to noise ratio in the SDSS $r$ band above 15. The
dataset comprises 459,953 spectra. In order to minimise the effect of
variations in the properties of the stellar populations on the
positions of the centroids of the spectral features, we have to select
a large number of targets -- so that the statistical variations are
evened out. Moreover, the sample has to comprise a galaxy population
as homogeneous as possible across the redshift window -- to reduce the
intrinsic scatter of the centroid positions {\sl per galaxy}. The
stellar velocity dispersion has been found to correlate strongly with
the properties of the stellar populations of galaxies
\citep{Bernardi:05}. We select galaxy spectra within a fixed interval
in velocity dispersion, $\sigma\in[100,250]$\,km/s (see
Fig.~\ref{fig:sample}). This velocity cut reduces our working sample
down to 329,867 galaxies. This selection ensures that we are typically
dealing with the same type of stellar populations.  However, in order
to study possible biases related to variations of the stellar content,
we explore in Sec.~\ref{sec:agemet} the effect of the properties of
the stellar populations in more detail, with a crucial result
presented in \S\S\ref{ssec:agemet2}. Furthermore, we analyze
sub-samples with respect to colour, velocity dispersion, equivalent
width of the lines, or signal-to-noise ratio (see Appendix).

\section{Methodology}
\label{sec:method}

\subsection{Spectral Line Selection}
\label{ssec:selection}

For each galaxy spectrum, we measure 60 prominent absorption or emission
features, and determine their central position. To
define the list of lines for the analysis, we build a reference
\"uber-spectrum by stacking 34,652 individual spectra within a
100--250\,km\,s$^{-1}$ range in velocity dispersion, and over a narrow
redshift interval (0.02$<$z$<$0.04). The stack is visually inspected
in order to select clean absorption (and some emission) lines. We
avoid lines that could not be easily isolated from neighbouring
features, and those with an asymmetric or flat
pattern. Fig.~\ref{fig:Lines} shows the stacked spectrum along with
the full set of 60 lines initially chosen for the analysis. We select
a narrow redshift bin as a reference, so that all spectral features
are directly calibrated {\sl with the same data}, avoiding biases caused by
the complexity of these features (as explained in the previous section).

It is important to note that the reference lines are not produced by
single atomic or molecular transitions. A galaxy spectrum is the
result of an unresolved superposition of a large number of stellar
spectra over a wide range of ages and chemical
composition. Furthermore, the dynamical state of the galaxy introduces
a strong Doppler broadening at the level of 50--400\,km\,s$^{-1}$, in
addition to the spectral resolution limit of the spectrograph
\citep[R=$\lambda/\delta\lambda\!\sim\!  2,000$ for the SDSS
  spectrograph, ][]{Smee:13}. Hence, each absorption line comprises a
blend of many spectral features from a large number of atomic, and
sometimes molecular species in the stellar atmospheres of the
population (i.e. a complex mixture of spectra from stars with
different mass, age, and chemical composition). This complication is
inherent to the intrinsic nature of galaxies and can not be overcome
by any further improvement in instrumentation.  Consequently, any
study such as the one we are pursuing here, i.e.  an attempt to
constrain departures from the standard cosmological redshift, would be
unfeasible by use of {\sl individual} galaxy spectra.

The large dataset provided by SDSS enables us to compare spectra from
hundreds of thousands of galaxies, averaging out the variations in
stellar properties. In addition, a careful selection of the galaxies
also minimises the variations expected from the underlying stellar
populations. In this sense, our adopted cut in velocity dispersion
homogenizes the galaxy population, and helps mitigate these potential
pitfalls. As the use of individual (i.e. atomic or molecular lines)
from laboratory measurements is unfeasible, our methodology rests on
making use of as high a number as possible of high quality galaxy
spectra, and a definition of {\sl effective central} wavelengths for a
number of features, measured directly on the galaxy spectra at some
fiducial redshift\footnote{In this paper we use the galaxies in our
  lower redshift bin to define the central position of the spectral
  lines as a reference. However, the results are not affected by this
  choice (see Appendix).}. With these effective values we can
accurately study {\sl relative} variations with respect to redshift
and wavelength.

The use of laboratory-measured lines or higher resolution stellar
spectra as absolute references is also impractical because of the
effect of residuals in the derivation of the index of refraction in
air.  The SDSS spectra are given with respect to vacuum wavelengths,
using the standard conversion adopted by the International
Astronomical Union \citep{Morton:91}.  Note this conversion quotes,
unchanged, an expression from an older paper
\citep{Edlen:53}. Residuals from the wavelength offset caused by a
different index of refraction of air between this reference and more
recent ones \citep{Ciddor:96} will give variations comparable to the
observed shifts. We note that the prescription of \citet{Ciddor:96} is
adopted as standard by the International Association of Geodesy and is
assumed to be more accurate for a wider range of conditions in
atmospheric temperarure and humidity. By defining as reference a large
number of galaxy spectra {\sl from the same sample}, within some
redshift bin (in our case the redshift range from z=0.02 to 0.04), we
self-calibrate the positions of the lines avoiding potential biases
from the air-to-vacuum wavelength conversion, and shifts of the
centroids of the features caused by complex line blending effects.
Additional biases due to the instrumental configuration should be
minimised as well.

\subsection{Determination of the centroids of spectral lines}
\label{ssec:centroids}

\begin{figure*}
\centering
\includegraphics[width=16cm]{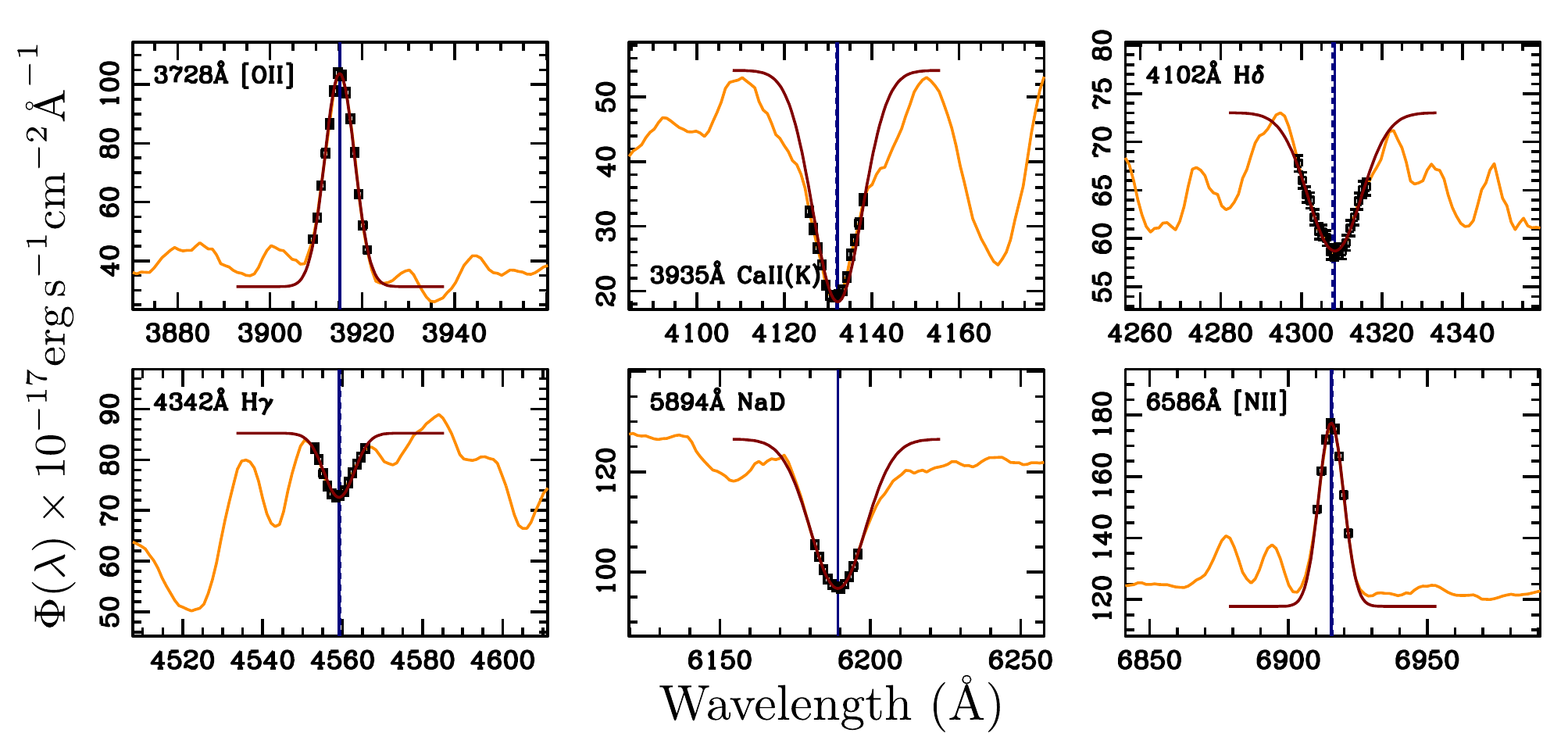}
\caption{Examples of the line fitting procedure. The figure shows an
  example of the line fitting process for a typical SDSS galaxy spectrum (ID:
  plate-mjd-fiber ID: 545-52202-166) at redshift z=0.0501 with
  velocity dispersion $\sigma=186$\,km\,s$^{-1}$ and signal-to-noise
  ratio in the SDSS-$r$ band of 35\AA$^{-1}$. The panels show 6 of the
  28 lines used in this analysis, as labelled.  The orange line is the
  actual spectrum; the data points used in the fit are shown with
  error bars. The best Gaussian fit is the red line. The vertical blue
  lines mark the retrieved central position of the spectral feature
  (solid) and the estimated position according to the redshift given
  by the SDSS standard pipeline (dashed).}
\label{fig:LineFit}
\vskip+0.15truein
\end{figure*}

The spectroscopic data are retrieved from the DR10 DAS server of the
Sky Digital Sky Survey\footnote{\tt
  http://skyserver.sdss.org/dr10/en/home.aspx}.  Only spectra are
retrieved with a signal-to-noise ratio (measured as a median value
within the SDSS-r passband) above 15.  The sample is further
restricted in redshift between 0.02 and 0.25. A dedicated code in C
reads the file, and performs line fitting on a number of prominent
lines. Each individual spectrum is continuum subtracted using the BMC
method of \citet{BMC}, with the standard choices of a 100\AA\ box
width and a 90\% level for the ``boosted median''.  This choice has
been comprehensively and independently tested to provide robust
estimates of the continuum in SDSS spectra \citep{Hawkins:14}. Each
line is subsequently fitted by a function
\begin{equation}
\label{eq:Fitfcn}
f(\lambda;{\cal C},{\cal A},\lambda_0,\sigma_\lambda)
= {\cal C} + {\cal A}e^{-\frac{(\lambda-\lambda_0)^2}{2\sigma_\lambda^2}}
\end{equation}
In order to remove spurious pixels, we mask those flagged as bad pixels
by the SDSS pipeline. In addition, we mask out pixels that are significantly
affected by airglow. We reject fits that use fewer than 6 data points for each
reference line.
The fitting procedure is based on the Levenberg-Marquardt \citep{NRC} algorithm to
fit the data to equation~\ref{eq:Fitfcn}, taking the measured fluxes within an interval
$\Delta\lambda$ centered at the expected position of the line. The
process is repeated several times for different choices of
$\Delta\lambda$, and the output corresponds to the one that 
gives the lowest value of the $\chi^2$ statistic, defined in the 
standard way, namely:
\begin{equation} 
\chi^2\equiv\sum_{i\in\Delta\lambda}\bigg[\frac{\Phi(\lambda_i)-f(\lambda_i;{\cal C},
{\cal A},\lambda_0,\sigma_\lambda)}{\sigma_i}\bigg]^2,
\label{eq:chi2}
\end{equation} 
where $\Phi(\lambda_i)$ and $\sigma_i$ are the observed flux and
uncertainty at the sampled wavelegths, respectively.  The output of
the code includes the line position ($\lambda_0$), 
amplitude (${\cal  A}$) and line width ($\sigma_\lambda$). It also gives out the
signal-to-noise ratio of the measurement, the equivalent width, and
the $\chi^2$ of the fit. Fig.~\ref{fig:LineFit} shows a typical
example of line fitting for a few of the spectral features targeted
in this analysis.

We note that the SDSS spectral pipeline (spec1d) provides for each
spectrum a block of data comprising fits of emission lines
(spZline\footnote{\tt https://www.sdss3.org/dr9/spectro/pipeline.php}).  A preliminary
exploration of this dataset showed that it is not good enough for the
purposes of this work, as the fitting procedure constrains sets of
lines to have the same redshift and the same width. Our method treats
each spectral feature separately, with independently derived central wavelengths
and line widths, and considers both emission and
absorption features.

\begin{figure*}
\centering
\includegraphics[width=8.4cm]{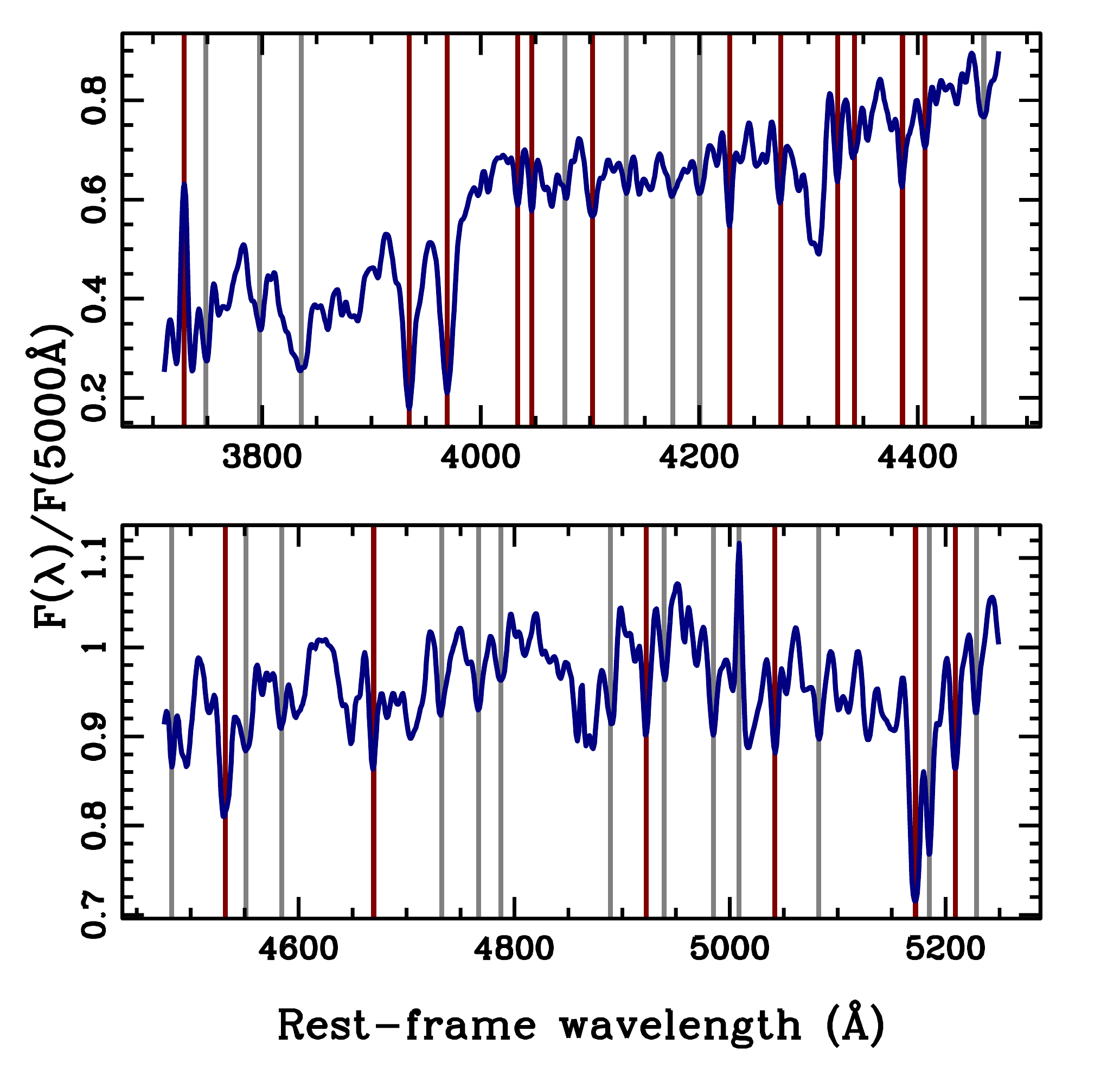}
\includegraphics[width=8.4cm]{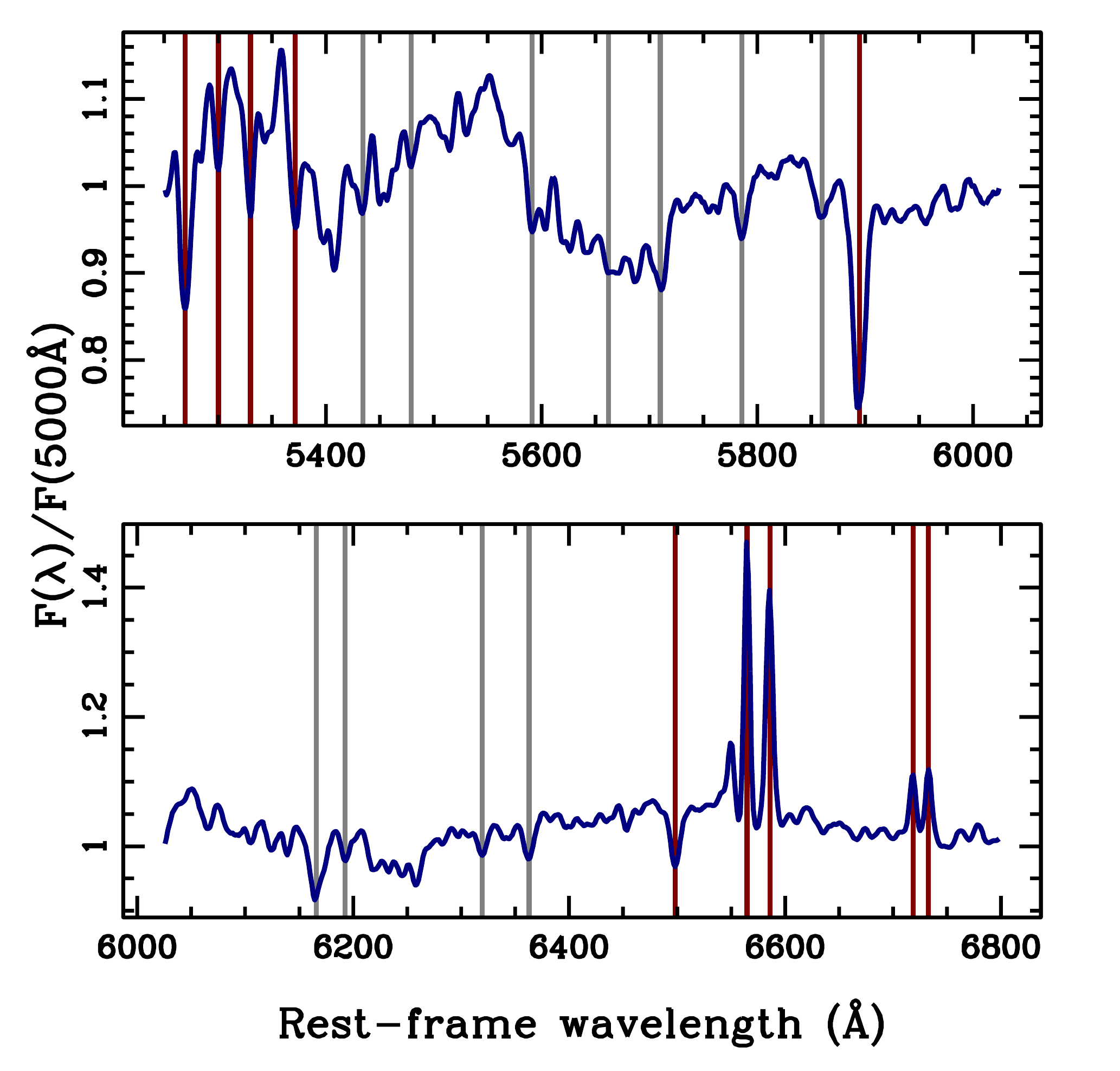}
\caption{Selection of spectral features. A stack of 34,652
  individual SDSS galaxy spectra from our sample over the redshift range
  z=0.02-0.04 is shown along with the position of the 60 targeted
  lines. The 28 lines selected for the final analysis are shown in red,
  the rejected lines appear in grey.}
\label{fig:Lines}
\vskip+0.15truein
\end{figure*}

We begin by running the code on all spectra within the reference
redshift bin (z=0.02-0.04), which comprises 34,652 galaxies. The
distribution of values of $\Delta(z,\lambda_0)$ for each line in this
redshift bin is studied. For each galaxy, z is taken as the median
value of the redshift from all the useful
lines. Fig.~\ref{fig:Hists} shows the distribution of
$\Delta(z,\lambda_0)$ for a few lines, showing the typical behaviour
of the whole set. The distributions can be split into two classes.
The top panels of the figure (blue) show the histograms for lines with
very small dispersion, whereas the bottom panels (red) show lines with
a much noisier behaviour.  For reference, the orange line corresponds
to a Gaussian distribution with $\sigma=100$\,km\,s$^{-1}$. The
difference between these two sets is caused by a complex mixture of
factors including the blending of neighbouring lines, the typical SNR
of the measurement, or the intrinsically higher variance of the
features. For our purposes, we take advantage of this comparison to
select a sub-sample of 28 {\sl bona-fide} spectral features
(Tab.~\ref{tab:LinesOK}, and red lines in Fig.~\ref{fig:Lines}),
rejecting the rest for the analysis (Tab.~\ref{tab:LinesNotOK}).  We
note that the extension of the analysis to the higher redshift bins
does not change the result. In other words, those lines that are well
behaved in the lowest redshift interval are also well behaved in the
higher redshift bins. In addition, these histograms allow us to
sharpen the definition of the centroids of each spectral
feature. Therefore, we re-define the central positions such that at
the reference redshift $\Delta(z_{\rm REF},\lambda_0)=0$ for all
lines. This restriction fixes the central wavelengths used as
reference when measuring the departure of the (1+z) law at higher
redshifts (see Tab.~\ref{tab:LinesOK}).  This relative calibration of
the line positions removes the biases described in previous sections.

As a comparison of our measurements with the official redshift
estimates from SDSS -- based on a cross-correlation with a set of
carefully defined templates \citep{SDSS:DR8} -- we find an offset
z$_{\rm SDSS}-\langle$z$_{\rm OURS}\rangle=(-0.981\pm 0.013)\times
10^{-5}$, where $\langle$z$_{\rm OURS}\rangle$, for each galaxy, is
given by the median of the redshift distribution measured by all
targeted lines.  Note that in contrast to the official SDSS redshift,
our methodology allows us to probe the redshift as a function of
wavelength. The RMS scatter of the difference between SDSS and our
redshift estimate is 22\,km\,s$^{-1}$, comparable to the quoted
uncertainty of {\sl individual} SDSS spectra \citep{SDSS:DR1}. To
explore any departure from the (1+z) law with redshift we divide
our sample in 15 redshift bins. As a simple rule of thumb, the number
of spectra within each of the 15 redshift bins is $\sim 20,000$, which
implies an accuracy per bin of 0.15\,km\,s$^{-1}$, or $\delta z/(1+z)\sim
5\times 10^{-7}$.

Fig.~\ref{fig:Dz} shows the results of our analysis, binned in four
rest-frame wavelength regions. Each bin includes a similar number of
spectral features. Note that in eq.~\ref{eq:Deltaz} we need to choose
a reference redshift for each galaxy. We opted for the median redshift
derived from all the spectral lines. The result does not change
significantly if the average is used instead. The error bars represent
the 68\% confidence level. In order to test whether the underlying
stellar populations are introducing a systematic in this result --
given that the spectral features themselves are blends of many lines
that can affect the centroid of the feature -- we also show as orange
and purple lines the results for sub-samples split according to
stellar velocity dispersion, an observable that strongly correlates
with the properties of the stellar populations
\citep{Bernardi:05}. Only a mild difference is found, suggesting that
the local properties of the galaxies can not explain the observed
departure of the (1+z) law.  In the Appendix, we show an extensive set
of tests on sub-samples split according to properties that could
introduce a systematic effect, such as the signal to noise ratio of
the measurement, the equivalent width of the spectral feature, the
observed colour of the galaxy, or the effect of telluric
absorption. We robustly obtain similar trends in all these tests.


\section{Analysis}
\label{sec:agemet}


\subsection{Testing the effect of the age and metallicity of the stellar population on the observed signal}
\label{ssec:agemet1}

Fig.~\ref{fig:Dz} shows a small but significant difference with
respect to velocity dispersion, where the galaxy spectra at low
velocity dispersion features the largest deviation from the standard
cosmological (1+z) law (i.e. the highest $|\Delta(z)|$). Although this
could be caused by the lower resolution produced with a higher stellar
velocity dispersion, one could question whether variations in the
properties of the underlying stellar populations are driving the trend
reported here.  Complex line blending effects affect the centroids of
the lines, and a systematic difference in the stellar population
composition with redshift could introduce a signal. To test this
potential bias, we select subsamples of galaxy spectra based on age-
and metallicity sensitive indicators.

We choose the Dn(4000) index of \citet{Dn4000} to trace the
age-sensitive 4000\AA\ break, and the [MgFe]$^\prime$ index defined by
\cite{Thomas:03} as a metallicity-sensitive indicator. We use the
index measurements of SDSS spectra from \citet{Brinch:04}, available
from the SDSS data server. The high SNR constraint imposed by our
selection criteria results in accurate index measurements on
individual galaxy spectra. The top panels of Fig~\ref{fig:Idx} shows
the index-index plot of the data in three redshift bins. Since
spectral indices are sensitive to velocity dispersion, we only use
galaxy spectra within the $\sigma\in[100,150]$\,km\,s$^{-1}$ interval.
The leftmost panel shows three tracks for a set of simple stellar
populations (SSPs) from the MILES-based models of \citet{MIUSCAT},
with a \citet{Kroup:01} IMF.  The SSPs for three different
metallicities: [Z/H]=$-$0.2 (blue), 0.0 (i.e.  solar, black) and
$+$0.2 (red). The tracks range in age from 0.1\,Gyr (bottom-left) to
13\,Gyr (top-right). For reference, three crosses mark the ages 1, 5
and 10\,Gyr for the track with solar metallicity. In this paper, we
only need to determine whether changes in the age and metallicity
distribution contribute to the observed departure from the standard
cosmological law. Therefore, we perform the same analysis for the
trend of $\Delta(z)$ with wavelength and redshift, as described above,
restricting the sample to subsets with markedly different properties
of age and metallicity.  The boxes on the three right-hand panels in
Fig.~\ref{fig:Idx} (top) represent the lowest and highest quartiles of
the distribution {\sl within each redshift bin}. These are the
subsamples we will use to explore this potential bias. We stress that
this subsample is restricted to velocity dispersions between 100 and
150\,km\,s$^{-1}$.  The bottom panels of Fig.~\ref{fig:Idx} show the
equivalent of Fig.~\ref{fig:Dz} for the young/metal-poor (Q1) and
old/metal-rich (Q4) subsamples. Notice we recover a similar separation
of the trend with redshift, as found with respect to velocity
dispersion. The highest departure, $\Delta(z)$, is found for the
younger component (i.e. the one with the lowest velocity dispersion).
However, the dominant trend remains unchanged. To further assess a
potential systematic caused by the effect of the stellar populations
on the line centroids, we explore in the next sub-section an
independent approach based on population synthesis modelling.


\subsection{Testing the effect of the methodology on the observed signal}
\label{ssec:agemet2}

The previous subsection, as well as the plethora of tests conducted in
the Appendix indicate that the observed trend in $\Delta(z,\lambda_0)$
can not be interpreted as a result of stellar population variations
with redshift. There remains the possibility that the trend is an
artifact of the methodology. To test this hypothesis, we resort to
population synthesis models.  We create sets of synthetic galaxy
spectra with the same sampling and spectral resolution as the SDSS
data.  We use the MILES-based models of \citet{MIUSCAT} with a
\citet{Kroup:01} IMF. Gaussian noise is added to the spectra, assuming
a SNR=30 in the SDSS-$r$ band window, typical of the observed
sample. Note the aim of this exercise is not to test the effect of SNR
on the result (see Appendix), therefore, we do not need to adopt the
same distribution of values of SNR. The data are redshifted according
to the standard (1+z) law, and the same methodology is applied to
these data to derive $\Delta(z)$. By construction, we should expect
$\Delta(z)=0$ in all redshift bins.

For simplicity, we adopt simple stellar populations, where the
metallicity is extracted from a Gaussian distribution with mean
[Z/H]=0.0 and RMS=0.2\,dex.  To assess the effect of different stellar
populations, we consider three different cases, with a formation
redshift fixed at z$_F=0.5$ (i.e. dealing with younger populations),
z$_F=2$ (older populations), or taking a uniform random number between
0.5 and 2 for this formation redshift, corresponding to a more complex
range of stellar ages.  For each of these three cases, the synthetic
sample comprises 10,000 mock spectra per redshift bin,
i.e. 3$\times$150,000 synthetic spectra in total.

\begin{figure*}
  \centering
  \includegraphics[width=16cm]{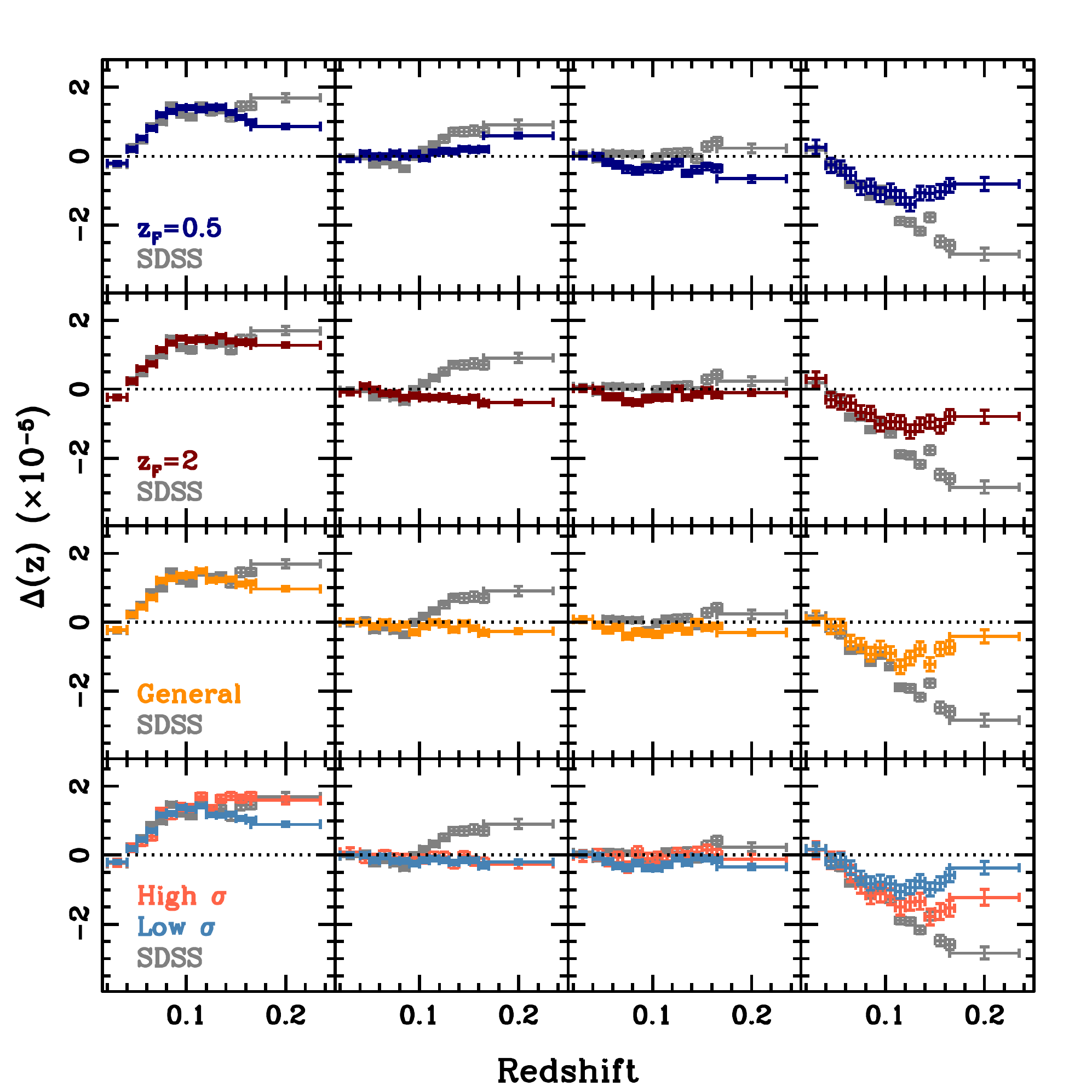}
  \caption{Trend of $\Delta(z)$ with redshift for a set of
    5$\times$150,000 mock galaxy spectra created from population
    synthesis models, with similar characteristics to the SDSS
    data. The coloured symbols correspond to samples where the
    formation redshift is fixed at $z_F=0.5$ (blue, top panels),
    representing overall younger populations; at $z_F=2$ (red, middle
    panels), for a family of older galaxies; or for a random set of
    values of $z_F$ between 0.5 and 2 (orange, bottom panels), which
    would better map the real populations. Note that a similar trend
    to the observed data (in grey) is produced by the synthetic
    samples, although this trend is relatively insensitive to the
    details of the underlying stellar populations. The top three
    panels are obtained from a sample with velocity dispersion $130\pm
    60$\,km\,s$^{-1}$. The bottom panel illustrates the effect of
    velocity dispersion, with two subsamples: $230\pm
    30$\,km\,s$^{-1}$ (high $\sigma$, red), and $130\pm
    30$\,km\,s$^{-1}$ (low $\sigma$, blue), see text for details.
 \label{fig:DzSynth}}
\vskip+0.15truein
\end{figure*}

The top three panels of Fig.~\ref{fig:DzSynth} show the trends of
$\Delta(z)$ derived for these three star formation histories, using a
sample with velocity dispersion drawn from a Gaussian distribution
with mean 130\,km\,s$^{-1}$ and RMS 60\,km\,s$^{-1}$.  For reference,
the grey symbols correspond to the observed trend in the SDSS galaxy
spectra. Although the observed result is not exactly reproduced, this
figure confirms that the variation in $\Delta(z)$ is caused by the
methodology used in this work. Before explaining how the methodology
introduces such an artifact, note the relatively small difference
between the three choices of formation epoch. Such an outcome suggests
the variation of $\Delta(z)$ does not depend strongly on the specific
details of the age and metallicity distribution of the stellar
component -- consistent with our negative findings in
\S\S\ref{ssec:agemet1}, or the subdominant effect when segregating the
sample with respect to velocity dispersion or color (see Appendix). On
the other hand, the velocity dispersion, that controls the width of
the spectral features, plays an important role on the strength of the
artificial signal.

\begin{figure*}
\centering
\includegraphics[width=16cm]{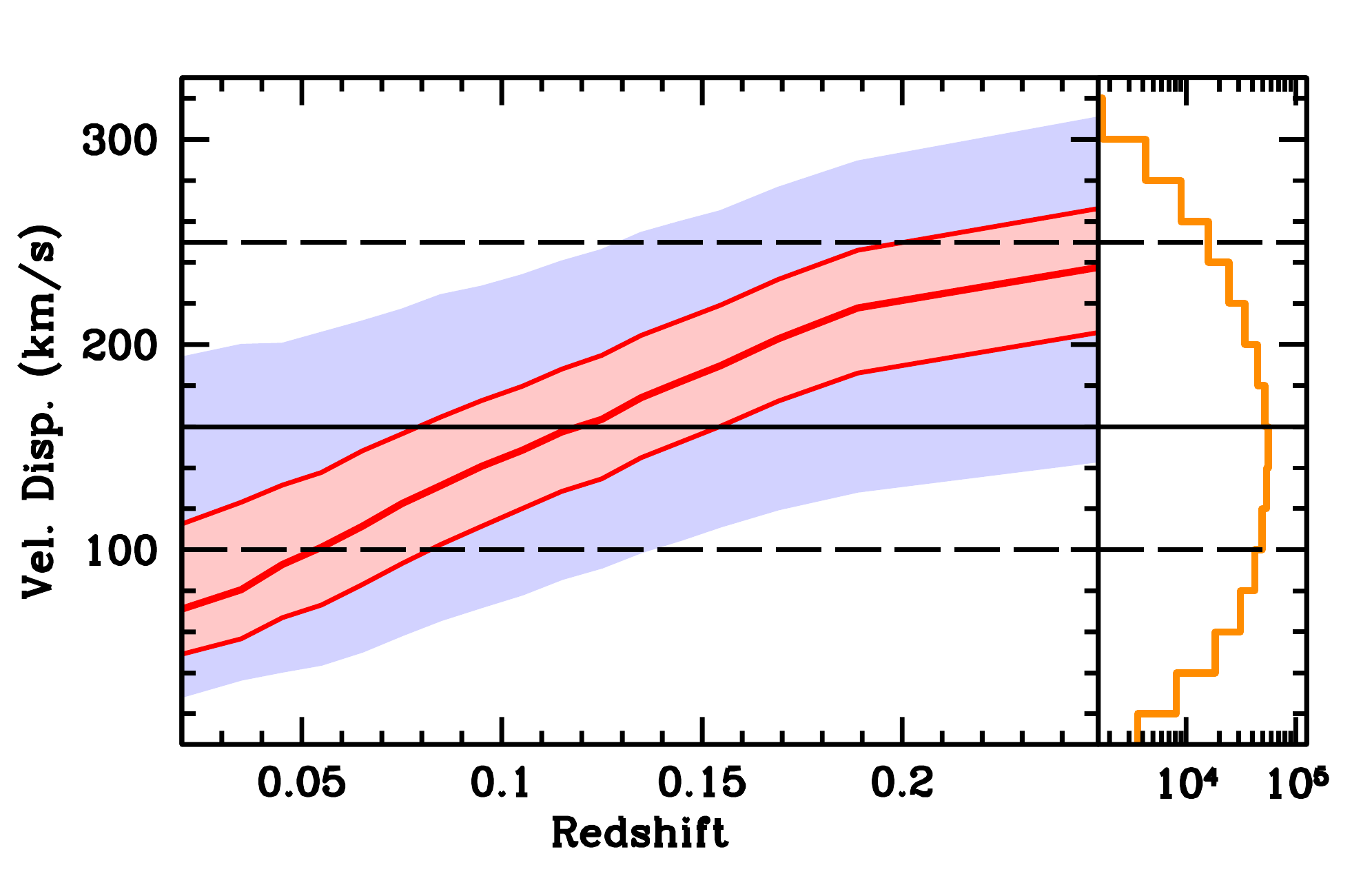}
\caption{The sample of SDSS spectra used in this analysis is shown on
  the left panel as a density distribution with respect to redshift
  and central velocity dispersion.  The thick red line traces the
  median of the distribution with respect to redshift, whereas the
  light red (light blue) shaded area delimits the region encompassing
  25-75\% (5-95\%) of the sample. The orange histogram on the right is
  the distribution of the full sample with respect to velocity
  dispersion. The horizontal dashed lines mark the limit enforced for
  the analysis of a homogeneous galaxy sample.}
\label{fig:sample}
\vskip+0.15truein
\end{figure*}

The lowest panel of Fig.~\ref{fig:DzSynth} shows the result for the
general case regarding the star formation history (i.e. a random
formation redshift between 0.5 and 2), but with two different velocity
dispersions: $230\pm 30$\,km\,s$^{-1}$ (high $\sigma$, in red) and
$130\pm 30$\,km\,s$^{-1}$ (low $\sigma$, in blue). Note the high
velocity dispersion sample gives more compatible results with the
observed SDSS data at high redshift (where the sample is dominated by
high velocity dispersion galaxies, see Fig.~\ref{fig:sample}).
Note that in the low redshift interval, z$\lesssim$0.1, the synthetic
data fully account for the signal. Therefore, in this case we can claim
that the cosmological (1+z) law holds to 1 part per million.

The results of this test confirm that the signal is produced by our
methodology. Moreover, the artifact is more significant in the sample
with higher velocity dispersion, for which the spectral features are
inherently wider. Why does our methodology introduce this signal? The
derivation of the line centroids relies on the minimization of the
$\chi^2$ given by Eq.~\ref{eq:chi2}. This minimization is conducted in
the observed wavelength space. Therefore, the average number of data
points for the fit increases with redshift. It is also important to
note that the number of points describing the spectral features also
depends on their width -- that is affected by the velocity
dispersion. The (1+z) stretching of the spectra also distorts the
shape of the lines with redshift. This implies that as the redshift
increases, the sampling of the lines in the observer frame will be
slightly denser on the red side of the line with respect to the blue
side. Hence, the red side of the lines have a higher effective weight,
with a resulting bias on the position of the centroids.  This effect
is expected to be more pronounced when the feature is wide and located
on the red side of the spectrum.  Therefore we conclude that the bulk
of the trend observed in SDSS spectra is explained as a change in
{\sl effective} resolution.

\section{Discussion}
\label{sec:discussion}

This paper presents a detailed analysis of a hypothetical departure of
the cosmological (1+z) law with respect to wavelength. By use of
hundreds of thousands of high-quality SDSS galaxy spectra, we target a
number of emission and absorption lines in the
3,700--6,800\AA\ rest-frame optical window, originally finding a
departure from this law, parameterised by $\Delta(z)$ (see
eq.~\ref{eq:Deltaz}), at the level of $\sim 5\times 10^{-5}$ over the
redshift range z$\lesssim 0.25$, noting that our statistical accuracy
reaches $\Delta\lambda/\lambda\sim 10^{-6}$.  Various systematic
effects are considered. We reach the final conclusion that the shapes
of the absorption lines -- adopted for the derivation of wavelength
references -- combined with the changing velocity dispersion of the
sample with respect to redshift, produces an effective change in
spectral resolution of these complex lines, creating the observed
trend in $\Delta(z)$.  No measurable departure from the standard (1+z)
law is detected to within one part in 100,000 in the general sample,
and to within one part per million in the redshift range
z$\lesssim$0.1.  As reference, if this result were associated to a
variable speed of light across cosmic time, the analysis presented
here would impose a change in $c$ between blue ($\lambda\sim$4,000\AA)
and red ($\lambda\sim$7,000\AA) optical photons of $\Delta
c_\lambda<$300\,m/s, out to z=0.1, corresponding to 1.3\,Gyr in cosmic
time.

The shape of the absorption lines originates from the velocity
dispersion of the stars moving under the gravitational potential of
the host galaxy.  The Doppler shift caused by this motion blends a
large number of atomic and molecular absorption lines, creating
effective absorption profiles whose centroids are affected in a
complex way when stretched according to redshift.  The typical values
of the stellar velocity dispersion in galaxies -- between $\sigma\sim
100$ and 300\,km\,s$^{-1}$ -- introduce an ``effective'' spectral
resolution R$\sim c/\sigma\sim 2,000$ that complicate the
determination of wavelength references. Interestingly enough, the
details of the age and metallicity distribution of the underlying
stellar populations play a subdominant role, as shown in
\S\ref{sec:agemet}, and the Appendix, whereas a simple model of
population synthesis -- that combine no more than $\sim 1,000$ stellar
spectra from the solar neighbourhood -- is capable of reproducing the
observed trend in galaxy spectra.

As an aside, our data, restricted to the optical window, cannot be
used to determine whether such an effect may be present in other
spectral regions.  Nonetheless, one can use the accurate measurement
of the blackbody spectrum of the cosmic microwave background (CMB), to
derive an upper limit to the expected maximum effect at longer
wavelengths.  The constraint provided by the CMB is not based on
spectral lines, but on the (excellent) fits to a blackbody
radiator. To provide a rough estimation on how the uncertainty in the
CMB temperature translates into an uncertainty on the equivalent of
$\Delta$(z), let us use the position of the peak of the CMB spectrum
as a wavelength reference. Using the current uncertainty on the
temperature of the CMB
\citep[T$_{\rm CMB}$=2.72548$\pm$0.00057\,K,][]{Fixsen:2009}, the
wavelength position of the peak is measured with the following
accuracy (applying Wien's law, taking the constant from
NIST\footnote{\tt http://physics.nist.gov/cgi-bin/cuu/Value?bwien}):

\begin{equation*}
\lambda_{\rm CMB,peak}=1063.22\pm0.43\ \mu m
\end{equation*}

Therefore the current uncertainty in the CMB allows us to measure the
effect with a precision $\Delta\lambda/\lambda\sim
4.04\times10^{-4}$. This is an order of magnitude {\sl larger} than
the limit met by our methodology in the optical window.

Within the optical regime, previous works aiming at measuring a
variable fine structure constant in quasar spectra have provided to
date the best constraints on wavelength departures from (1+z), with an
upper limit at the level of
\citep[see, e.g.,][]{Webb:01,Quast:04,Rahmani:14,Albareti:15}
$\delta\lambda_z/[(1+z)\delta\lambda_0]\sim 10^{-5}-10^{-6}$ out to
redshifts $z\sim 1$, where $\delta\lambda$ denotes the separation
between two spectral lines, and the subscript refers to the
redshift. Our results are not in contradiction with this upper
limit. The studies of the variation of the fine structure constant use
multiplets of a given atomic element. The targeted spectral features,
such as the [OIII] doublet ($\lambda\lambda$4960, 5008\AA), cover a
relative narrow spectral window ($\Delta\lambda$$\lesssim$100\AA). In
comparison, our study covers a wavelength window at least 30 times
larger.

In this context, we wonder whether analyses of the variation of the
fine structure constant based on absorption lines
\citep[e.g.,][]{Quast:04} may suffer from similar type of systematic
effects. In addition, the use of nebular emission lines might present
equivalent systematic complications in the determination of the
centroid of the line, this time caused by the gas kinematics, but such
musings are beyond the scope of this paper.

Is the result shown in this paper an absolute limit to the accuracy
one can achieve with galaxy spectra? Although the present data cannot
provide an accuracy better than one part in 100,000 out to
z$\lesssim$0.25, we speculate that high SNR spectra at significantly
higher resolution -- $R\gtrsim 10,000$ -- can be used to probe in
detail the effective shapes of the absorption lines, potentially
allowing us to break this barrier. The denser sampling at higher
resolution will decrease the statistical weight in the methodology due
to differences in the number of fitting points between the blue and
the red sides of the spectral features, caused by the stretching of
the spectra.

\acknowledgments

The authors would like to thank Juan Betancort Rijo, St\'ephane
Courteau, Jos\'e Alberto Rubi\~no Mart\'\i n, Fernando Atrio
Barandela, Francesco La Barbera, Mart\'\i n L\'opez Corredoira, and
Alexandre Vazdekis for insightful comments about this project.  The
referee, Dr Dan Kelson, is warmly thanked for his positive criticism,
suggestions, and very hard work! I.F. acknowledges support to visit
the IAC through the Severo Ochoa visitor programme. I.T. acknowledges
support by the ``Programa Nacional de Astronom\'\i a y Astrof\'\i
sica'' of the Spanish Ministry of Science and Innovation under grant
AYA2013-48226-C3-1-P. Funding for SDSS-III has been provided by the
Alfred P. Sloan Foundation, the Participating Institutions, the
National Science Foundation, and the U.S. Department of Energy Office
of Science. The SDSS-III web site is http://www.sdss3.org/.


\begin{figure*}
\centering
\includegraphics[width=16cm]{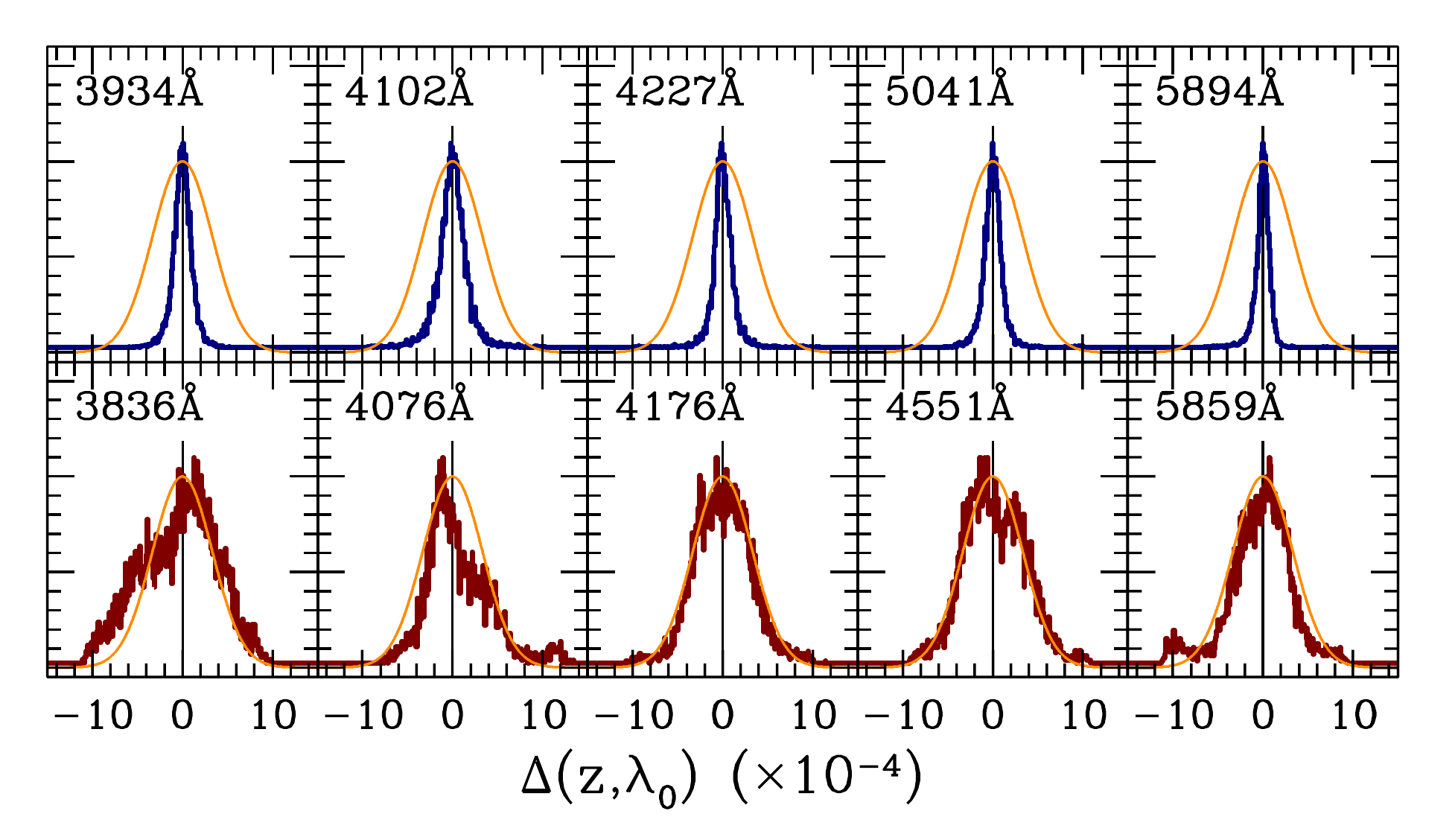}
\caption{Behaviour of $\Delta(z,\lambda_0)$ for individual
  lines. Typical examples are shown for lines either included ({\sl
    top}), or rejected ({\sl bottom}) in the final analysis. The
  histograms show the extracted value of $\Delta(z,\lambda_0)$ (as
  defined in eq.~1).  Each panel includes the position of the central
  wavelength in the rest-frame.  The orange line corresponds to a
  Gaussian distribution with $\sigma=100$\,km\,s$^{-1}$.}
\label{fig:Hists}
\vskip+0.15truein
\end{figure*}

\begin{figure*}
  \centering
\includegraphics[width=16cm]{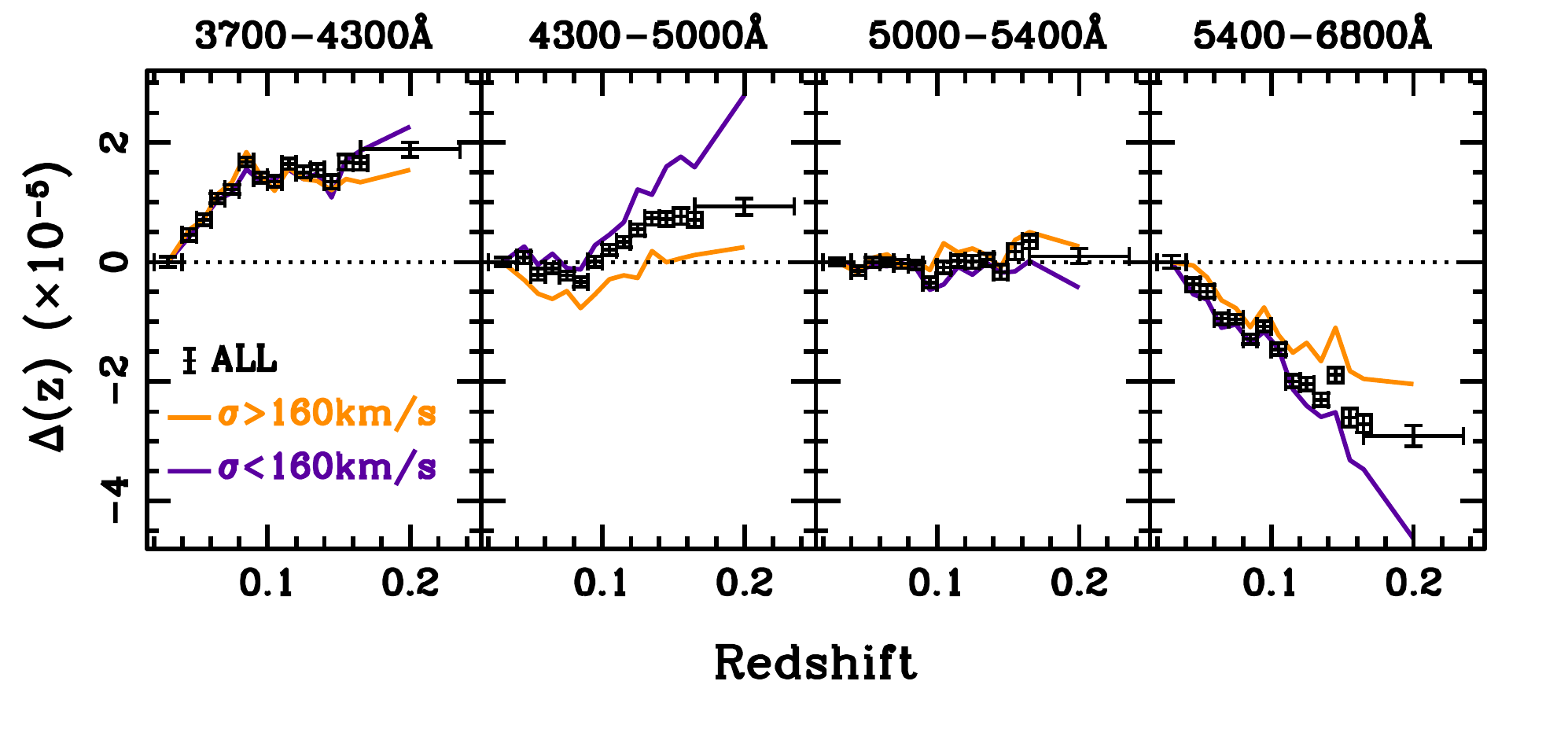}
\caption{Deviation of the galaxy spectra with respect to a standard,
  wavelength-independent cosmological law, given as
  $\Delta(z,\lambda_0)$ (see eq.~\ref{eq:Deltaz}; the standard
  case corresponds to $\Delta(z,\lambda_0)=0$).  All the
  information from the spectral features is binned into four
  wavelength regions, as labelled, chosen to keep a similar number of
  spectral lines per wavelength bin. The data is binned in 15 steps
  between redshift z=0.02 and z=0.25. The error bars mark the mean and
  its uncertainty. The orange and purple lines correspond to
  subsamples split with respect to velocity dispersion, as labelled.}
 \label{fig:Dz}
\vskip+0.15truein
\end{figure*}

\begin{figure*}
  \centering
  \includegraphics[width=16cm]{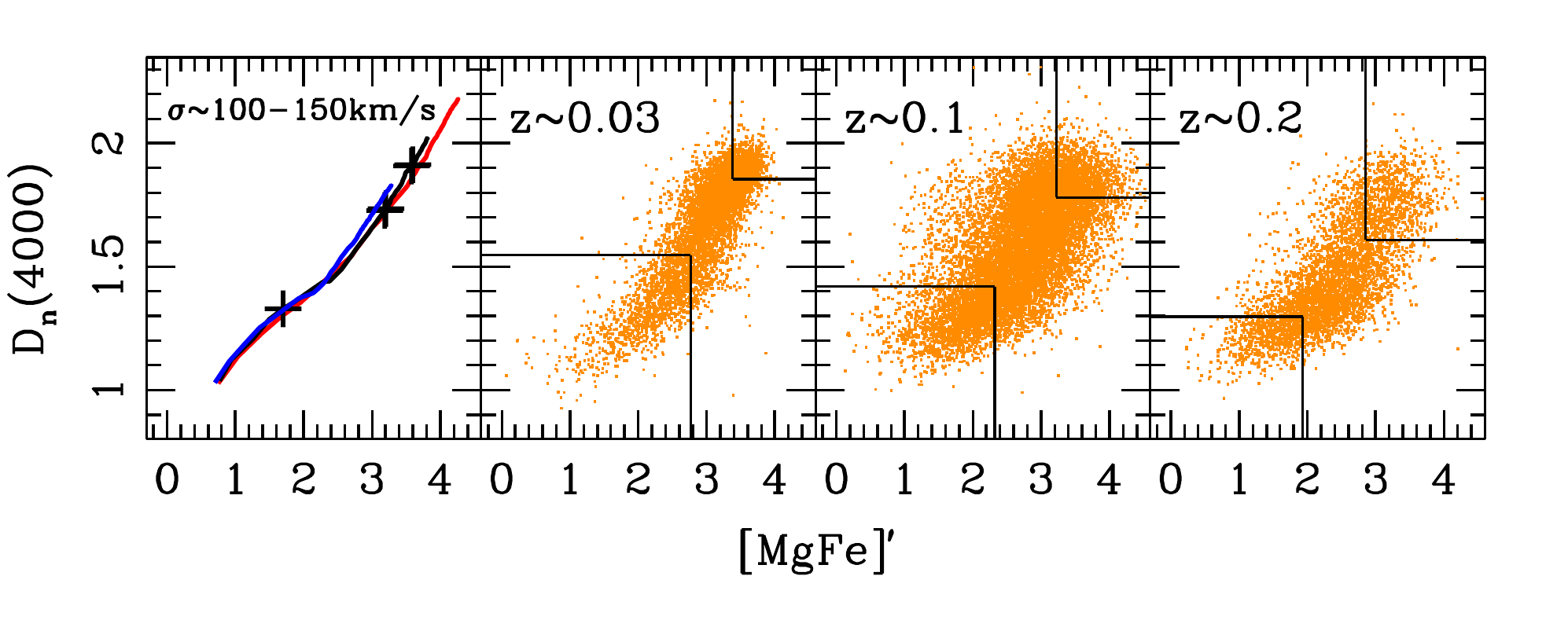}\\
  \includegraphics[width=16cm]{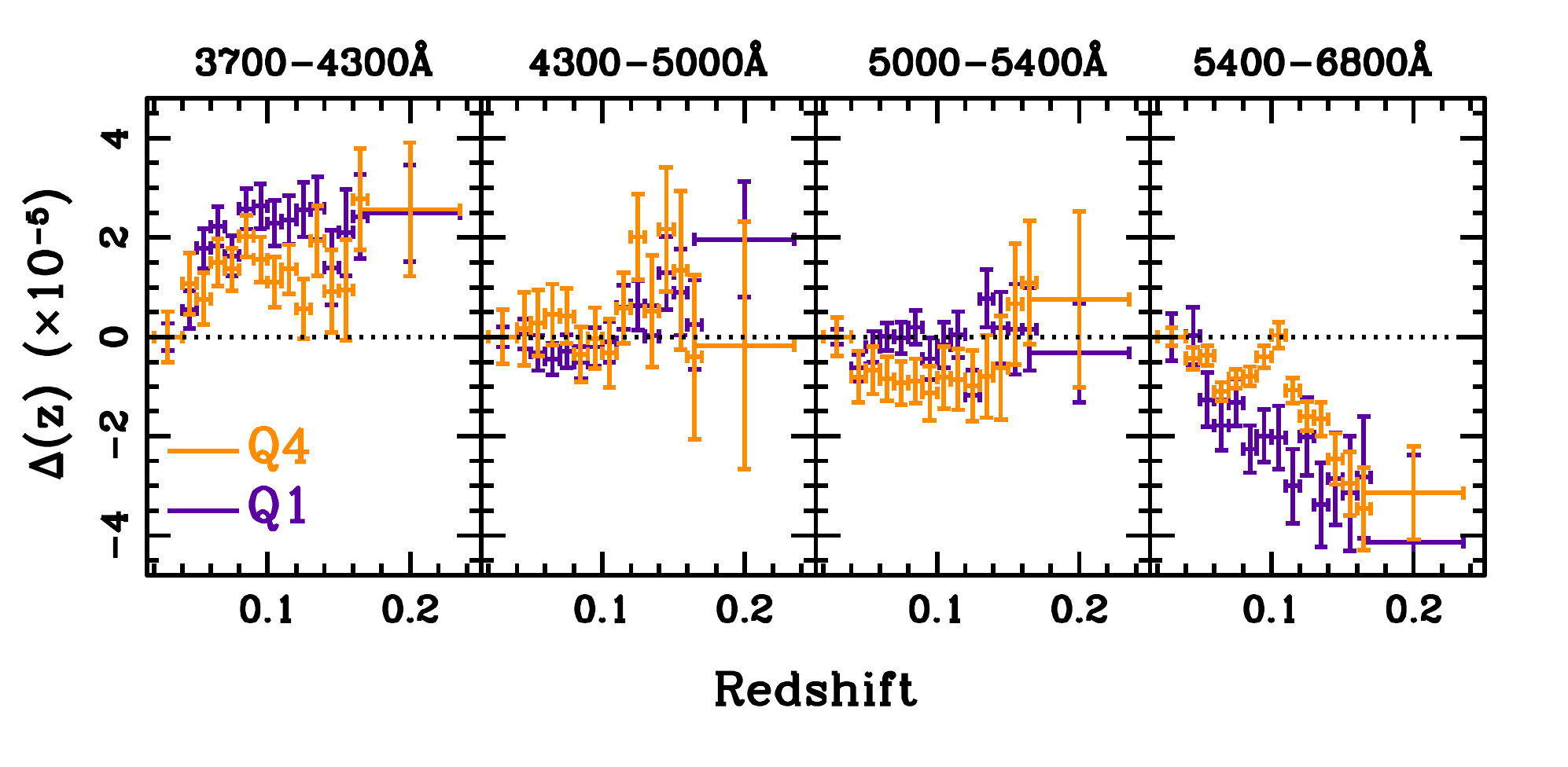}\\
  \caption{{\sl Top:} Selection of subsamples based on the age and
    metallicity of the underlying stellar populations (restricted to a
    velocity dispersion between 100 and 150\,km\,s$^{-1}$). The
    horizontal axis shows a metal-sensitive index ([MgFe]$^\prime$),
    and the vertical axis corresponds to the age-sensitive
    4000\AA\ break (see text for details).  The leftmost panel shows
    the result for a set of simple stellar populations with
    metallicities [Z/H]=-0.2 (blue), 0.0 (black) and +0.2 (red), over
    an age range from 0.1\,Gyr (bottom-left) to 13\,Gyr
    (top-right). For reference. crosses mark the positions at 1, 5,
    and 10\,Gyr for the model with solar metallicity. The three panels
    on the right show the characteristic distribution of index values
    for three redshift bins from our sample. The boxes mark the lowest
    and highest quartiles of the distribution.  {\sl Bottom:} Results
    of the trend of $\Delta(z)$ with redshift restricted within each
    redshift bin to the lowest quartile (Q1: young and metal-poor
    populations, purple), and highest quartile (Q4: old and metal-rich
    populations, orange).}
 \label{fig:Idx}
\vskip+0.15truein
\end{figure*}

\clearpage


\begin{deluxetable*}{cccccc}
\tabletypesize{\scriptsize}
\tablecaption{List of accepted spectral features (list restricted to
  the reference redshift bin: z=0.02-0.04). Col.~1 gives
  the identification number of the line; col.~2 is the central wavelength
  position {\sl after re-defining the position using the z=0.02-0.04
    sample as reference}; col.~3 is the number of galaxy spectra used
  to redefine the position of the central wavelength; col.~4 and 5
  give the average and RMS scatter of the measured equivalent width
  and signal-to-noise ratio, respectively. Col.~6 identifies the
  spectral feature either by a single atomic transition, or by a
  well-defined spectral index commonly used in stellar population
  spectroscopy \citep{Lick}.\label{tab:LinesOK}}
\tablewidth{14cm}
\tablehead{
\colhead{Line} & \colhead{Wavelength} &  \colhead{n} & 
\colhead{EW}  & \colhead{SNR} & \colhead{Comment}\\
\colhead{ID} & \colhead{\AA}  & \colhead{ }  & \colhead{\AA} & \colhead{ }
& \colhead{ }\\
\colhead{(1)} & \colhead{(2)} &  \colhead{(3)} & 
\colhead{(4)}  & \colhead{(5)} & \colhead{(6)}
}
\startdata
 00  & $3728.3714$ & $ 7287$ &  $-9.57\pm 47.18$ &  $ 31.81\pm  16.64$ & [OII]\\
 04  & $3934.7674$ & $21265$ &  $ 3.26\pm  1.18$ &  $ 28.36\pm  15.39$ & CaII (K)\\
 05  & $3969.3108$ & $21035$ &  $ 3.48\pm  1.61$ &  $ 32.48\pm  17.55$ & CaII (H)+H$\epsilon$\\
 06  & $4033.7667$ & $12439$ &  $ 0.69\pm  0.45$ &  $ 59.84\pm  26.79$ & FeI\\
 07  & $4046.7865$ & $11113$ &  $ 0.66\pm  0.47$ &  $ 56.95\pm  25.28$ & FeI\\
 09  & $4102.2947$ & $14503$ &  $ 1.16\pm  0.50$ &  $ 61.43\pm  27.16$ & H$\delta$\\
 13  & $4227.7748$ & $12615$ &  $ 1.00\pm  0.48$ &  $ 62.85\pm  27.43$ & CaI\\
 14  & $4274.2991$ & $15636$ &  $ 0.95\pm  0.40$ &  $ 65.60\pm  27.68$ & FeI\\
 15  & $4326.5288$ & $13817$ &  $ 1.07\pm  0.42$ &  $ 73.18\pm  31.19$ & FeI/CH\\
 16  & $4341.7585$ & $12545$ &  $ 1.16\pm  0.45$ &  $ 76.12\pm  31.21$ & H$\gamma$\\
 17  & $4385.9679$ & $13459$ &  $ 0.83\pm  0.35$ &  $ 74.26\pm  33.10$ & Fe4383\\
 18  & $4406.4512$ & $17680$ &  $ 0.70\pm  0.33$ &  $ 79.89\pm  32.32$ & FeI\\
 21  & $4532.1095$ & $17730$ &  $ 0.89\pm  0.44$ &  $ 95.65\pm  37.36$ & Fe4531\\
 24  & $4669.4250$ & $18104$ &  $ 0.60\pm  0.24$ &  $ 89.18\pm  33.42$ & C$_2$4668/FeII/ScII\\
 29  & $4922.0688$ & $19491$ &  $ 0.65\pm  0.22$ &  $ 96.43\pm  35.69$ & FeII\\
 33  & $5041.5147$ & $19601$ &  $ 0.54\pm  0.21$ &  $100.88\pm  37.71$ & FeI/SiII\\
 35  & $5171.8832$ & $22770$ &  $ 1.18\pm  0.41$ &  $ 91.01\pm  33.21$ & Mgb/MgI\\
 37  & $5209.2082$ & $20042$ &  $ 0.69\pm  0.23$ &  $102.18\pm  38.80$ & CrI\\
 39  & $5269.5436$ & $22210$ &  $ 1.00\pm  0.47$ &  $104.62\pm  39.48$ & Fe5270\\
 40  & $5300.1071$ & $20124$ &  $ 0.54\pm  0.23$ &  $109.68\pm  39.50$ & CrI\\
 41  & $5329.6268$ & $21427$ &  $ 0.63\pm  0.23$ &  $105.84\pm  39.01$ & Fe5335\\
 42  & $5371.5712$ & $18110$ &  $ 0.48\pm  0.23$ &  $114.78\pm  41.59$ & FeI\\
 50  & $5894.4307$ & $20871$ &  $ 1.13\pm  0.61$ &  $127.02\pm  49.20$ & NaD\\
 55  & $6498.2663$ & $21521$ &  $ 0.52\pm  0.20$ &  $145.62\pm  44.21$ & BaII/TiI/FeI/CaI\\
 56  & $6564.7901$ & $18729$ &  $-4.93\pm  6.77$ &  $154.33\pm  34.90$ & H$\alpha$\\
 57  & $6585.7938$ & $20563$ &  $-2.99\pm  3.10$ &  $151.57\pm  39.02$ & [NII]\\
 58  & $6718.6739$ & $18619$ &  $-1.67\pm  2.16$ &  $141.27\pm  39.37$ & [SII]\\
 59  & $6732.8674$ & $18175$ &  $-1.43\pm  1.60$ &  $140.97\pm  40.01$ & [SII]\\
\enddata
\end{deluxetable*}

\begin{deluxetable*}{cccccc}
\tabletypesize{\scriptsize} 
\vskip+1truein
\tablecaption{List of rejected spectral features (list restricted to
  the reference redshift bin: z=0.02-0.04). The identification of
  the columns follows Tab.~\ref{tab:LinesOK}.\label{tab:LinesNotOK}
}
\tablewidth{14cm}
\tablehead{
\colhead{Line} & \colhead{Wavelength} &  \colhead{n} & 
\colhead{EW}  & \colhead{SNR} & \colhead{Comment}\\
\colhead{ID} & \colhead{\AA}  & \colhead{ }  & \colhead{\AA} & \colhead{ }
& \colhead{ }\\
\colhead{(1)} & \colhead{(2)} &  \colhead{(3)} & 
\colhead{(4)}  & \colhead{(5)} & \colhead{(6)}
}
\startdata
 01  & $3748.4949$ & $15272$ &  $ 1.96\pm  0.99$ &  $ 25.29\pm  12.65$ & FeI\\ 
 02  & $3797.7169$ & $13623$ &  $ 1.68\pm  0.80$ &  $ 31.37\pm  16.51$ & FeI\\ 
 03  & $3835.8680$ & $13614$ &  $ 1.57\pm  1.18$ &  $ 27.88\pm  14.31$ & FeI\\ 
 08  & $4076.9786$ & $10166$ &  $ 0.49\pm  0.42$ &  $ 56.95\pm  25.82$ & FeI\\ 
 10  & $4133.0955$ & $12271$ &  $ 0.52\pm  0.38$ &  $ 64.66\pm  27.21$ & FeI\\ 
 11  & $4175.8954$ & $14534$ &  $ 0.54\pm  0.38$ &  $ 68.34\pm  29.50$ & FeI/FeII\\ 
 12  & $4200.4144$ & $15634$ &  $ 0.69\pm  0.37$ &  $ 68.64\pm  28.20$ & CH/FeI\\ 
 19  & $4460.3677$ & $16024$ &  $ 0.68\pm  0.38$ &  $ 85.95\pm  34.84$ & CaI/MnI/FeI\\ 
 20  & $4482.4370$ & $ 9568$ &  $ 0.18\pm  0.34$ &  $ 78.74\pm  33.13$ & MgII\\ 
 22  & $4551.0929$ & $13401$ &  $ 0.45\pm  0.32$ &  $ 90.13\pm  36.16$ & FeII/TiII\\ 
 23  & $4583.9470$ & $15040$ &  $ 0.38\pm  0.24$ &  $ 91.97\pm  35.90$ & FeII/CrII\\ 
 25  & $4732.5682$ & $15838$ &  $ 0.44\pm  0.25$ &  $ 98.07\pm  38.59$ & FeI\\ 
 26  & $4766.6875$ & $17808$ &  $ 0.37\pm  0.22$ &  $ 99.33\pm  37.71$ & FeI\\ 
 27  & $4787.3854$ & $15123$ &  $ 0.41\pm  0.25$ &  $104.52\pm  39.96$ & MnI\\ 
 28  & $4888.9960$ & $12538$ &  $ 0.62\pm  0.27$ &  $ 98.50\pm  38.50$ & FeI\\ 
 30  & $4939.1298$ & $17205$ &  $ 0.50\pm  0.23$ &  $103.22\pm  37.17$ & FeI\\ 
 31  & $4984.7983$ & $18919$ &  $ 0.52\pm  0.23$ &  $100.29\pm  36.70$ & Fe5015/FeI/TiI\\ 
 32  & $5008.2542$ & $ 6706$ &  $-2.19\pm  2.71$ &  $101.72\pm  34.35$ & [OIII]\\ 
 34  & $5082.4862$ & $18281$ &  $ 0.42\pm  0.20$ &  $102.48\pm  36.99$ & FeI\\ 
 36  & $5184.9476$ & $14656$ &  $ 0.58\pm  0.32$ &  $ 87.27\pm  31.28$ & Mgb/MgI\\ 
 38  & $5228.6874$ & $17714$ &  $ 0.46\pm  0.20$ &  $107.07\pm  40.76$ & FeI\\ 
 43  & $5433.8488$ & $15634$ &  $ 0.40\pm  0.21$ &  $109.89\pm  40.98$ & FeI\\ 
 44  & $5478.9656$ & $19052$ &  $ 0.33\pm  0.20$ &  $112.36\pm  42.43$ & FeI\\ 
 45  & $5591.1070$ & $ 5586$ &  $ 0.34\pm  0.17$ &  $112.41\pm  41.27$ & FeI/CaI/NiI\\ 
 46  & $5661.9125$ & $  474$ &  $ 0.27\pm  0.21$ &  $115.14\pm  46.53$ & FeI/ScII\\ 
 47  & $5710.1112$ & $ 4201$ &  $ 0.37\pm  0.23$ &  $121.29\pm  52.21$ & Fe5709\\ 
 48  & $5785.6318$ & $18512$ &  $ 0.36\pm  0.21$ &  $133.79\pm  55.23$ & Fe5782\\ 
 49  & $5859.8918$ & $15794$ &  $ 0.23\pm  0.19$ &  $137.94\pm  54.83$ & CaI\\ 
 51  & $6166.0155$ & $17983$ &  $ 0.37\pm  0.22$ &  $141.05\pm  44.56$ & CaI\\ 
 52  & $6192.7507$ & $18713$ &  $ 0.23\pm  0.15$ &  $141.46\pm  43.24$ & CaI\\ 
 53  & $6319.4290$ & $18845$ &  $ 0.26\pm  0.16$ &  $140.20\pm  43.22$ & FeI\\ 
 54  & $6362.8491$ & $19588$ &  $ 0.26\pm  0.23$ &  $142.72\pm  43.20$ & CaI\\ 
\enddata
\end{deluxetable*}

\clearpage

\appendix

\section{Potential Systematics}
\label{sec:systematics}

\subsection{Wavelength Calibration}
\label{ssec:wcalib}

\subsubsection{Testing the wavelength calibration of SDSS spectra}

\begin{figure*}
  \centering
 \includegraphics[width=5.4cm]{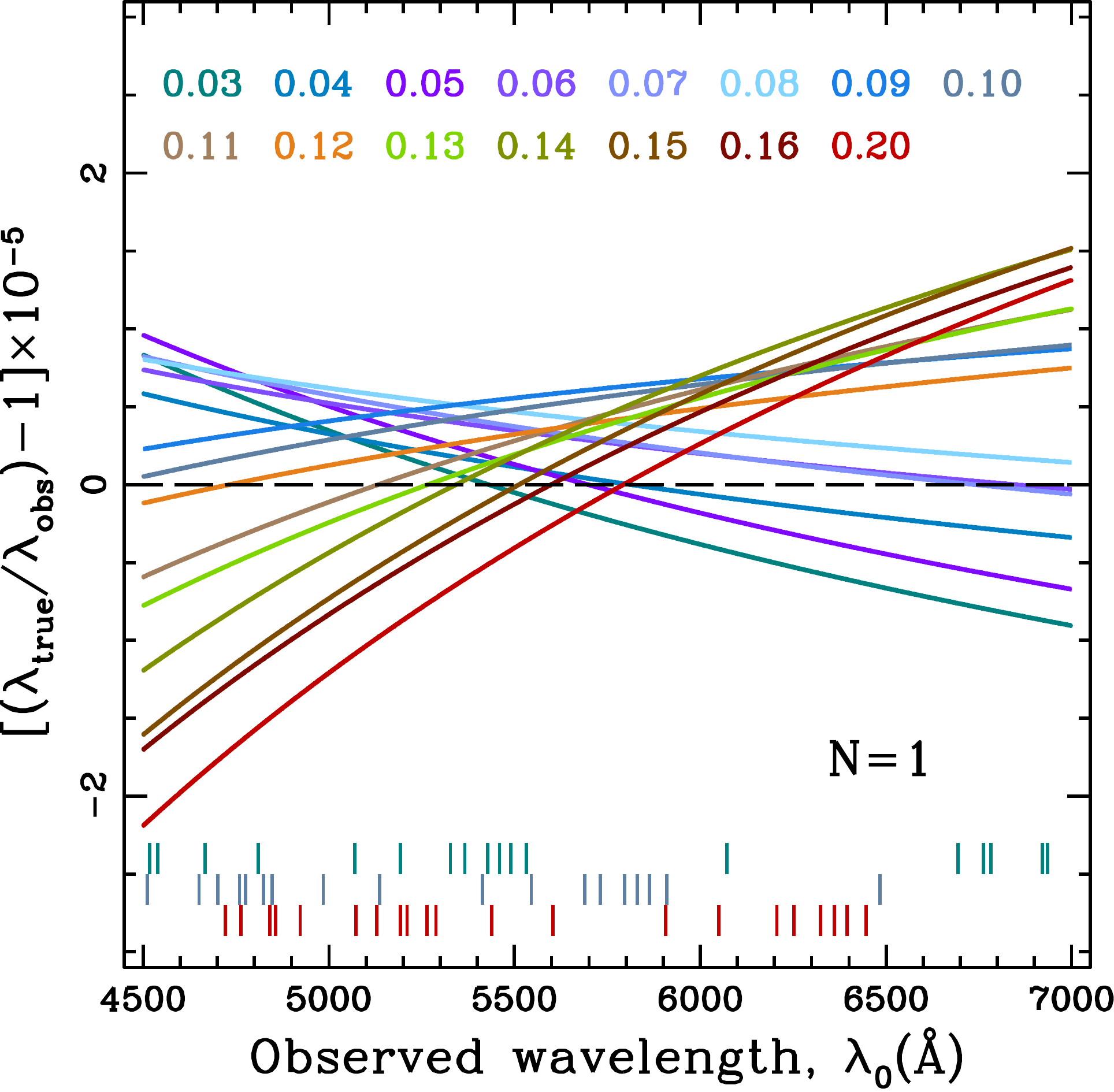}
 \includegraphics[width=5.4cm]{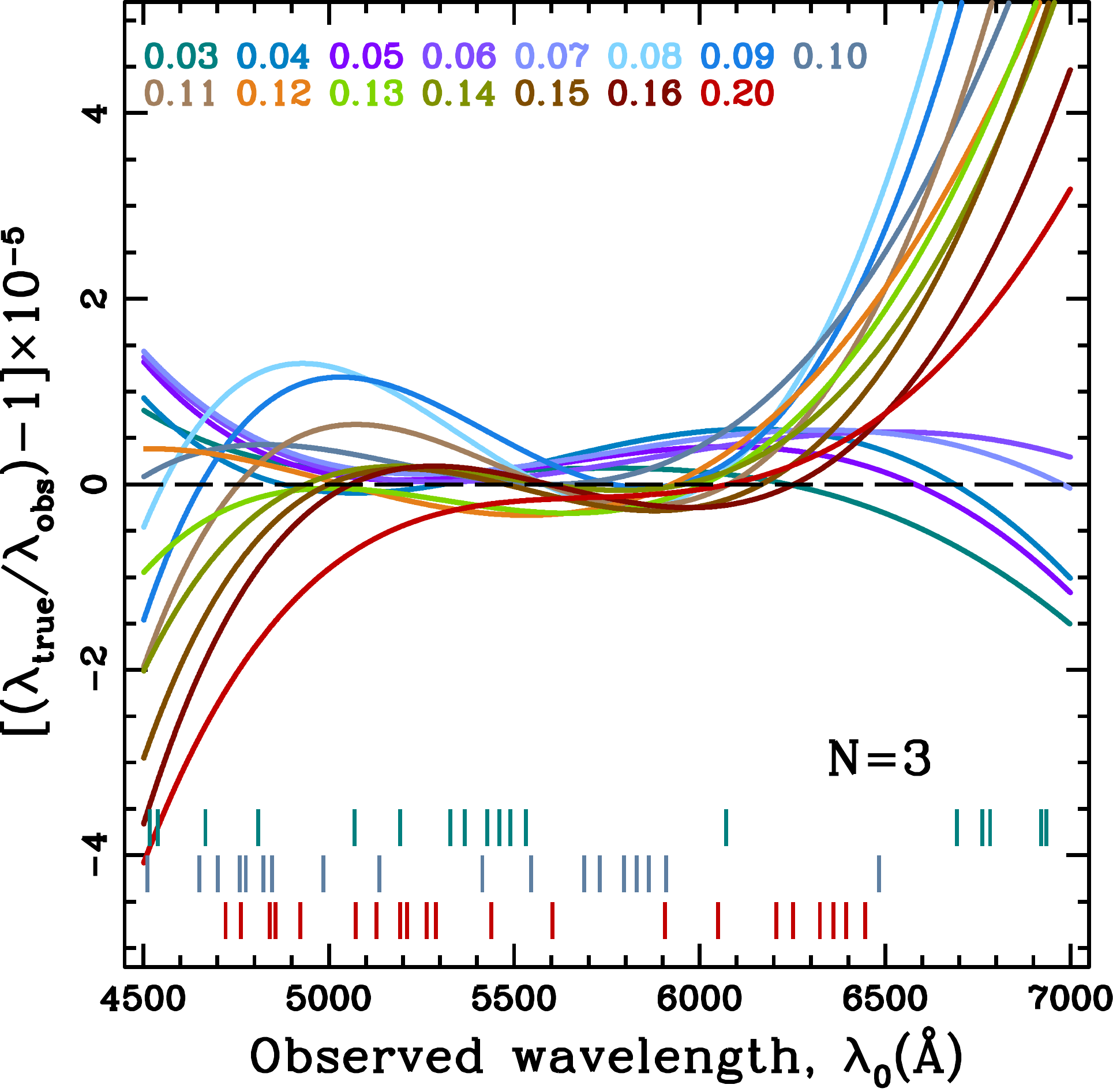}
 \includegraphics[width=5.4cm]{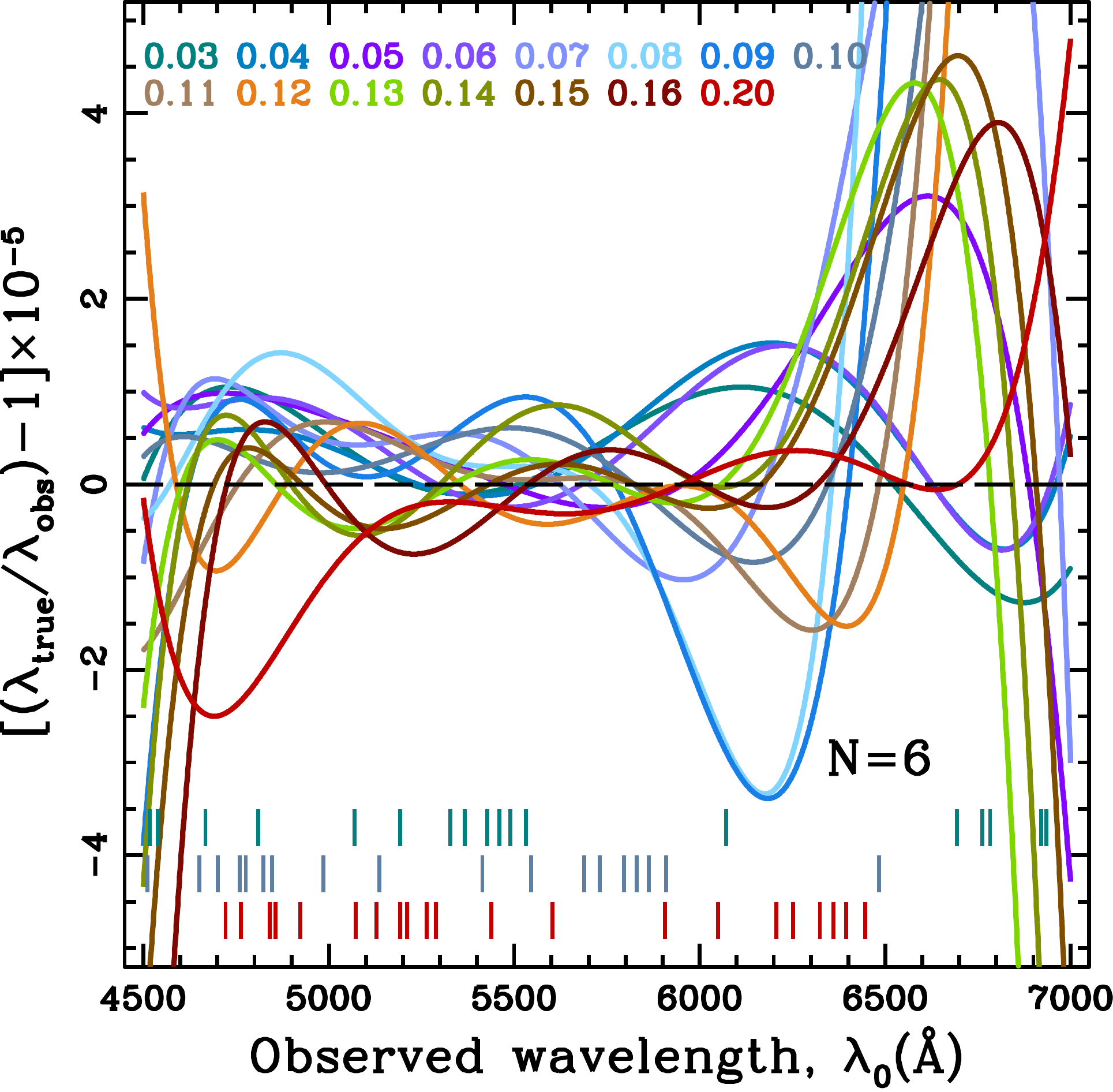}
  \caption{Testing a potential wavelength calibration residual. The
    most important systematic in our analysis is the effect of a
    residual in the wavelength calibration of the SDSS data. This
    figure tests such a scenario by adjusting to several polynomial
    transformations that recover the standard (1+z) cosmological law,
    with N=1 ({\sl left}); N=3 ({\sl middle}); N=6 ({\sl right}).
    Note that only the observer-frame wavelength range in common
    with  all galaxy spectra at all redshifts is used here. The hypothesis of a
    calibration residual is rejected as each redshift (coloured lines
    with each label representing the average redshift in each bin)
    gives different solutions, with a manifest trend with redshift.
    The marks in the lower side of the panels show the position of the
    lines included in the analysis for three of the redshifts
    considered, from top to bottom: z=0.03,0.10,0.20.}
 \label{fig:waveCal}
\vskip+0.15truein
\end{figure*}

In addition to the signal found in the simulated data (Sec.~\ref{sec:agemet})
one could also consider the possibility of an inaccurate wavelength calibration of the
SDSS spectra.  The official SDSS pipeline quotes typical errors
\citep{SDSS:DR7} of order 2\,km\,s$^{-1}$, which translates into an
uncertainty of $\delta\Delta(z;\lambda_0)\sim 7\times 10^{-6}$ for
individual spectra. Although this effect should be significantly lower
than our measurements, especially if we consider that the whole
analysis rests on the determination of approximately 6.8 million
individual spectral features, we nevertheless explore below a possible
systematic from a wavelength calibration trend. If the observed signal
is due to a calibration wavelength problem, then we should not expect
the calibration bias to depend on the redshift bin
considered. Consequently, we conduct the following test: for each
redshift bin we select only those lines within the common
observer-frame window, between $4500$ and $6800$\AA. We calculate in
this region how much the observed wavelengths ($\lambda_{\rm obs}$)
depart from the standard (1+z) law for each redshift bin. In this
test, we hypothesize that such departures are caused by an inaccurate
wavelength calibration, so that the actual measurement of the
wavelength should be $\lambda_{\rm true}=\lambda_0(1+z)$. We model
this wavelength calibration mismatch using a polynomial relation with
$\lambda_{\rm obs}$ in each redshift bin,
\begin{equation}
\lambda_{\rm TRUE}=\sum_{i=0}^N a_i\lambda_{\rm OBS}^i=\lambda_0(1+z).
\end{equation}
Fig.~\ref{fig:waveCal} shows the result for three choices of N: a
linear fit (N=1; {\sl left}); a cubic polynomial (N=3; {\sl middle})
or a sixth-order polynomial (N=6; {\sl right}).  If a wavelength
calibration were to explain the observed trend, we should find 
no significant variation of the solution with respect to redshift --
as we use in this test the same set of spectral lines in the observer
frame. The variation of the solutions with redshift confirms that the
calibration solution is not compatible among redshift
bins. Furthermore, this correction evolves in a smooth way with
redshift, illustrating the fact that this result cannot be
explained by a wavelength calibration residual.

\subsubsection{Independent wavelength calibration test with sky lines}

\begin{figure*}
  \centering
  \includegraphics[width=12cm]{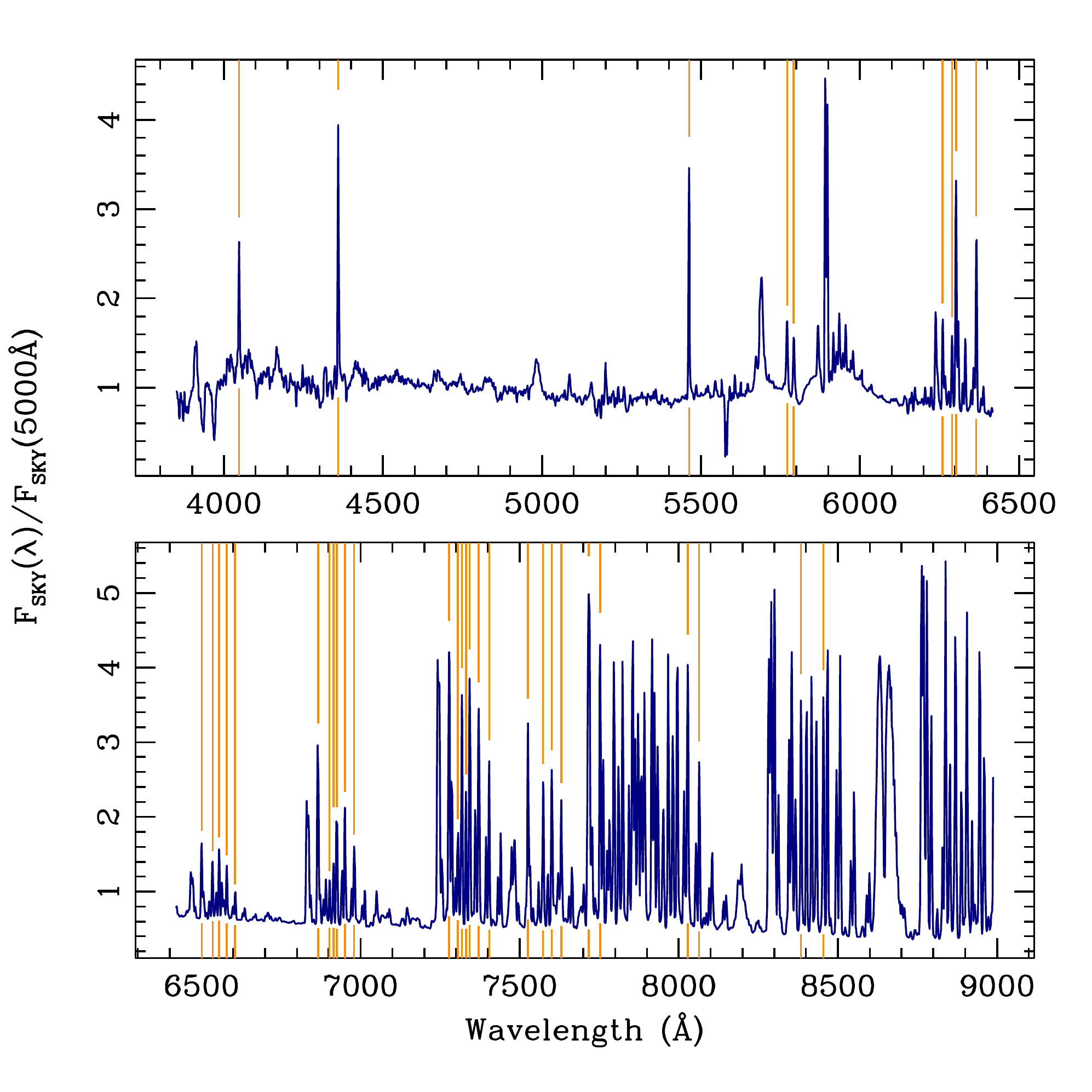}
  \caption{Stacked spectrum of the night sky obtained by median
    combining 34,652 SDSS spectra from the reference sample (i.e.
    corresponding to galaxies with z=0.02-0.04). The positions of
    several emission lines used as reference for an independent test
  of the wavelength calibration is shown as orange vertical lines.}
 \label{fig:Sky_stack}
\vskip+0.15truein
\end{figure*}

\begin{figure*}
  \centering
  \includegraphics[width=15cm]{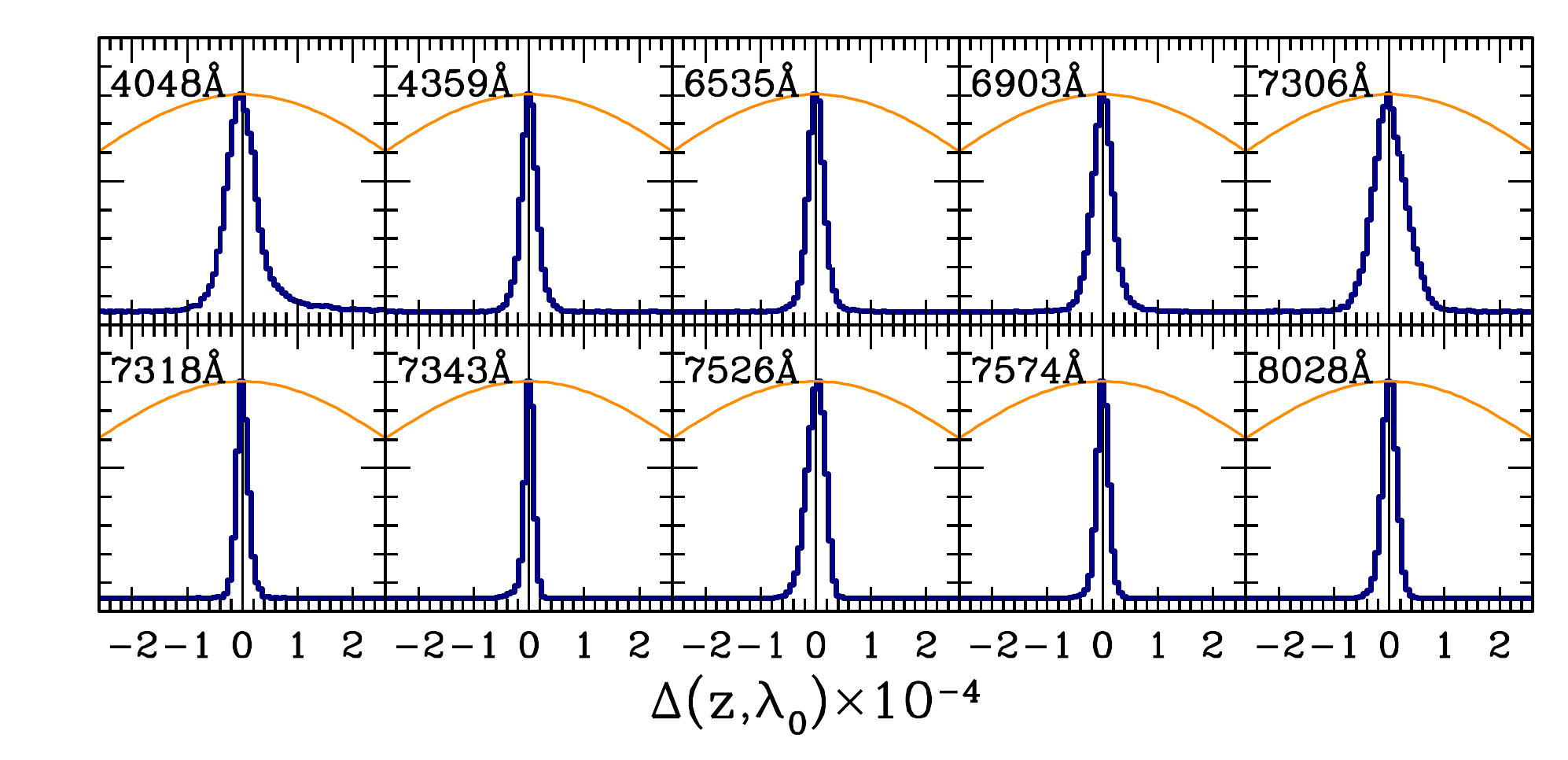}
  \includegraphics[width=15cm]{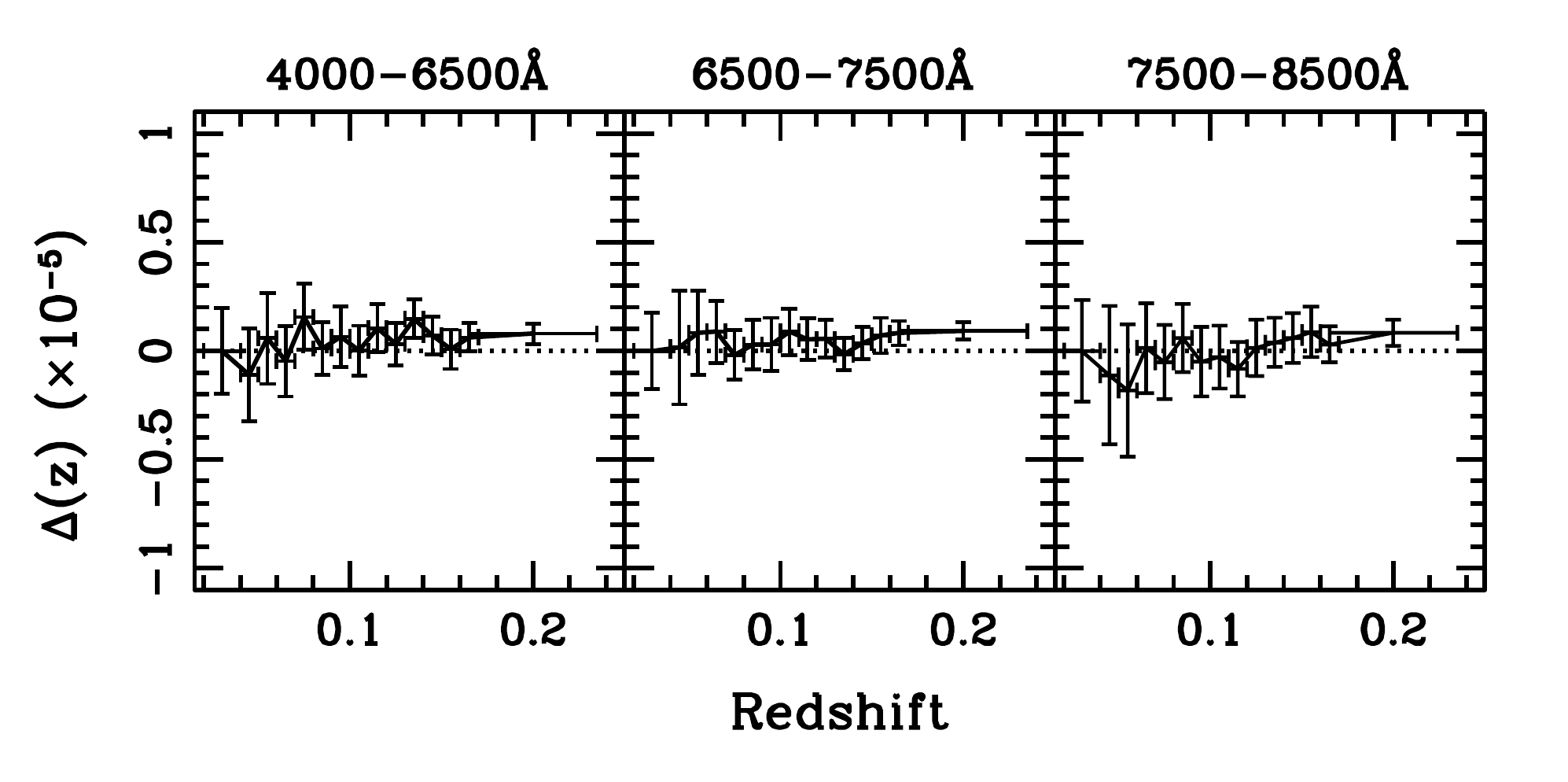}
  \caption{{\sl Top:} The equivalent of
      Fig.~\ref{fig:Hists} for a set of emission lines from the night
      sky, used as an additional test of the wavelength
      calibration of the SDSS data. The orange line corresponds to a Gaussian
      distribution with $\sigma=100$\,km\,s$^{-1}$.
      {\sl Bottom:} This figure presents the equivalent of $\Delta(z)$ measured
      {\sl directly on the night sky lines} in each of the galaxy spectra
      (hence the association of a redshift).}
 \label{fig:Sky_Hist}
\vskip+0.15truein
\end{figure*}

\begin{figure*}
  \centering
  \includegraphics[width=13cm]{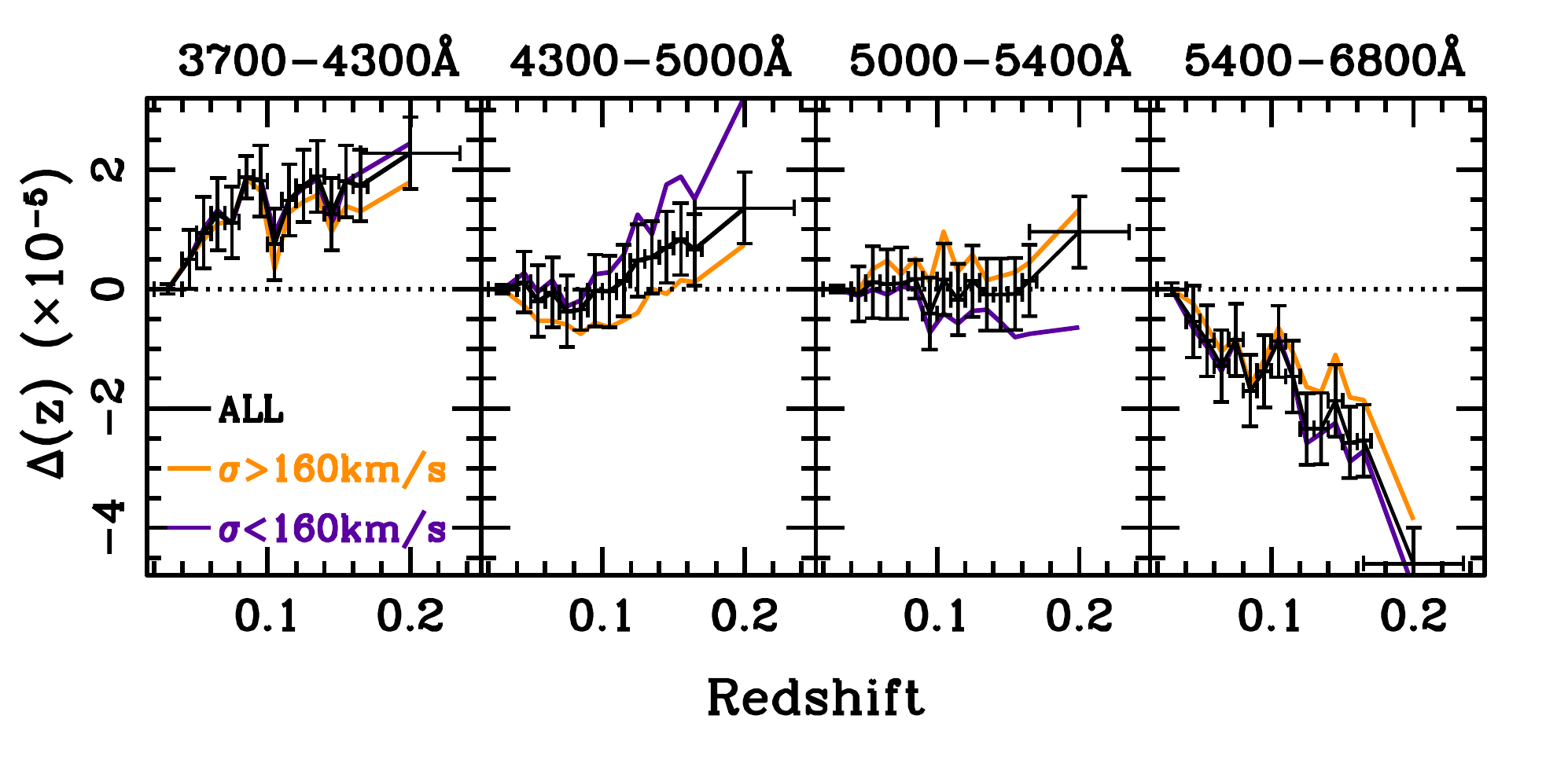}
  \caption{The equivalent of Fig.~\ref{fig:Dz}, using 
    the positions of night sky lines as reference, instead of the standard
    wavelength calibration solution. Note the vertical axis covers the
  same interval in $\Delta(z)$ as in Fig.~\ref{fig:Dz}.}
 \label{fig:Dz_Sky}
\vskip+0.15truein
\end{figure*}

As a further test of one of the most important
  instrumental systematic in this analysis, we consider a method where the
  derivation of the redshift offsets, $\Delta(z,\lambda_0)$, is, to lowest
  order, independent of the wavelength calibration solution. In this
  case, we make use of the sky spectra available for each SDSS galaxy
  spectrum. The positions of several night sky lines are used as
  reference points, interpolating between them to derive the positions
  of the spectral lines in the galaxy spectra. Therefore, this
  approach compares a set of wavelengths measured in galaxy spectra,
  $\{\lambda_i\}$ -- that vary according to the redshift of the galaxy
  -- and lines in sky spectra whose central positions, $\{\mu_i\}$, are constant
  throughout the sample. Let us assume that a non-zero wavelength
  calibration residual is present, such that the observed measurements
  and the true measurements can be written:
  \begin{equation*}
    \lambda_i^{\rm TRUE}=\lambda_i^{\rm OBS}\Big[1+f(\lambda_i^{\rm OBS})\Big].
  \end{equation*}
  Given that our findings reveal a small effect with respect to the
  standard (1+z) law, we can safely assume that $f\ll 1$. The same
  relation applies to the wavelength positions of the sky lines, $\{\mu_j\}$, namely,
  \begin{equation*}
    \mu_j^{\rm TRUE}=\mu_j^{\rm OBS}\Big[1+f(\mu_j^{\rm OBS})\Big]
    \Rightarrow 1+f(\mu_j^{\rm OBS})=\left(\frac{\mu_j^{\rm TRUE}}{\mu_j^{\rm OBS}}\right)
  \end{equation*}
  Therefore, the wavelengths measured in the galaxy can be written
  with respect to the sky wavelengths as:
  \begin{equation}
    \lambda_i^{\rm TRUE}\simeq\lambda_i^{\rm OBS}\left(\frac{\mu_j^{\rm TRUE}}{\mu_j^{\rm OBS}}\right),
  \end{equation}
  where the correction term on the right hand side is taken for the
  closest sky line associated to each galaxy line
  (i.e. $\lambda_i^{\rm OBS}\sim \mu_j^{\rm OBS}$), and we assume the
  wavelength positions of the sky lines in galaxies within the
  reference redshift bin (z=0.02--0.04), are the true wavelength
  values for night sky lines. In a nutshell, we are using the {\sl
    relative} offsets between the galaxy lines and the sky lines to
  derive the redshift of individual features, with the assumption that
  the positions of the night sky lines do not change systematically
  with respect to the redshift of the galaxy observed.
  \\
  The retrieved SDSS data include the night sky spectrum as an additional entry
  in each of the FITS files.  In order to define a list of strong sky
  lines, we build a stack sky spectrum by combining 34,652 individual
  ones from the sample within the reference redshift
  bin. Fig.~\ref{fig:Sky_stack} shows the stack along with the lines
  selected visually for the analysis. We selected 37 lines, making
  sure the SNR was high enough in individual sky spectra, and the line
  was not affected by the presence of neighbouring lines. The top
  panels of Fig.~\ref{fig:Sky_Hist} show the equivalent of
  Fig.~\ref{fig:Hists} for several sky lines. Note these histograms
  are much narrower, reflecting the simpler behaviour of the sky lines
  with respect to the blended features observed in galaxy spectra. The
  orange line, for reference, is also a Gaussian corresponding to a
  velocity dispersion of 100\,km\,s$^{-1}$. The bottom panels of
  Fig.~\ref{fig:Sky_Hist} show the accuracy of the wavelength
  calibration. We show the $\Delta(z)$ values as a function of
  redshift -- as in Fig.~\ref{fig:Dz} but estimated {\sl directly from
    the measurements of the sky lines}. Therefore, the redshift
  for these measurements corresponds to that of the galaxy spectrum
  from which the centroids of the night sky lines are measured.
  A zero value of $\Delta(z,\lambda_0)$ is expected throughout. The
  figure confirms the lack of a systematic trend, and illustrates the
  level of accuracy of our dataset, namely
  $\Delta\lambda/\lambda\sim 10^{-6}$.
  \\ 
  Fig.~\ref{fig:Dz_Sky} is the equivalent of Fig.~\ref{fig:Dz} when
  the redshifts are derived using the position of the night sky lines
  as a reference, instead of the standard wavelength calibration
  solution.  Note that the trend is consistent with the results using
  the standard wavelength calibration solution, therefore our results
  cannot be explained by a residual in this calibration.

\subsection{Asymmetry of the spectral features}
\label{ssec:asymmetry}

\begin{figure*}
\centering
\includegraphics[width=13cm]{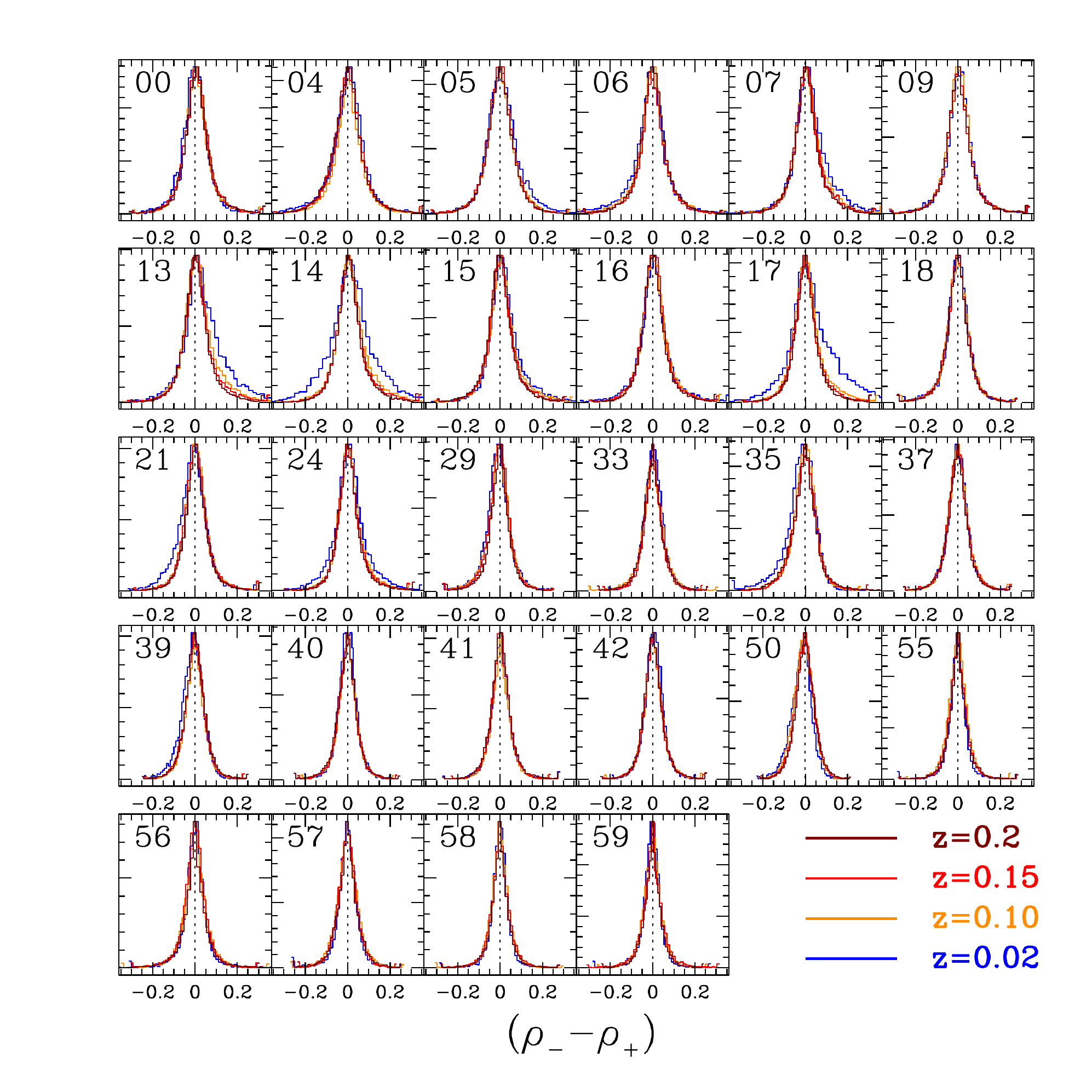}\\
\caption{Distribution of the difference between
    residuals on the blue and the red side of the lines (see
    Tab.~\ref{tab:LinesOK}) for several redshift bins. An asymmetric
    line profile may induce offsets in the derivation of the central
    position of the line. This figure shows that although some small
    variations are present in some of the lines, they cannot produce a
    cumulative effect that would mimick the trend presented in
    Fig.~\ref{fig:Dz}.}
\label{fig:Skew}
\vskip+0.15truein
\end{figure*}

\begin{figure*}
  \centering
  \includegraphics[width=14cm]{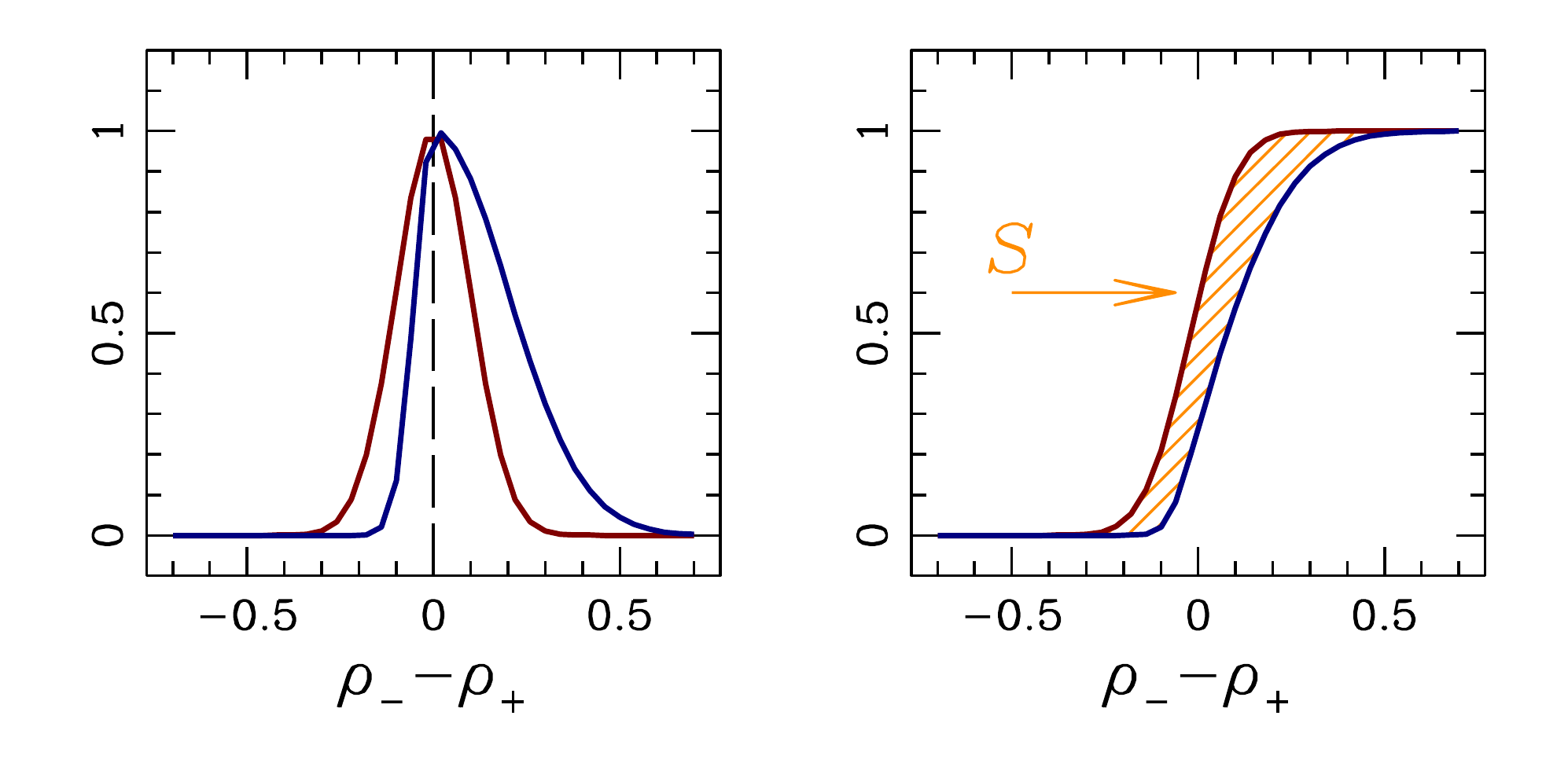}
  \includegraphics[width=14cm]{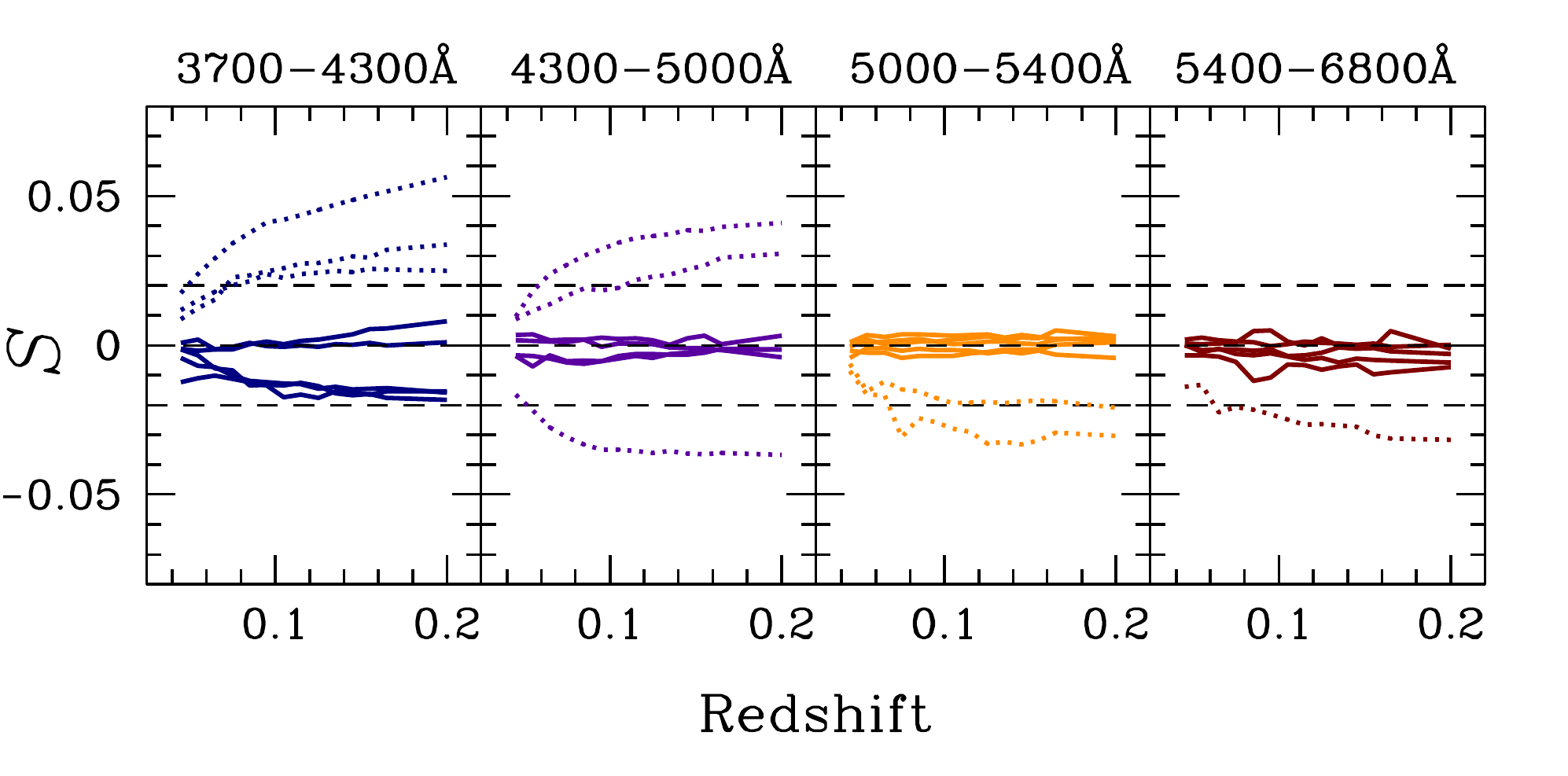}
  \includegraphics[width=14cm]{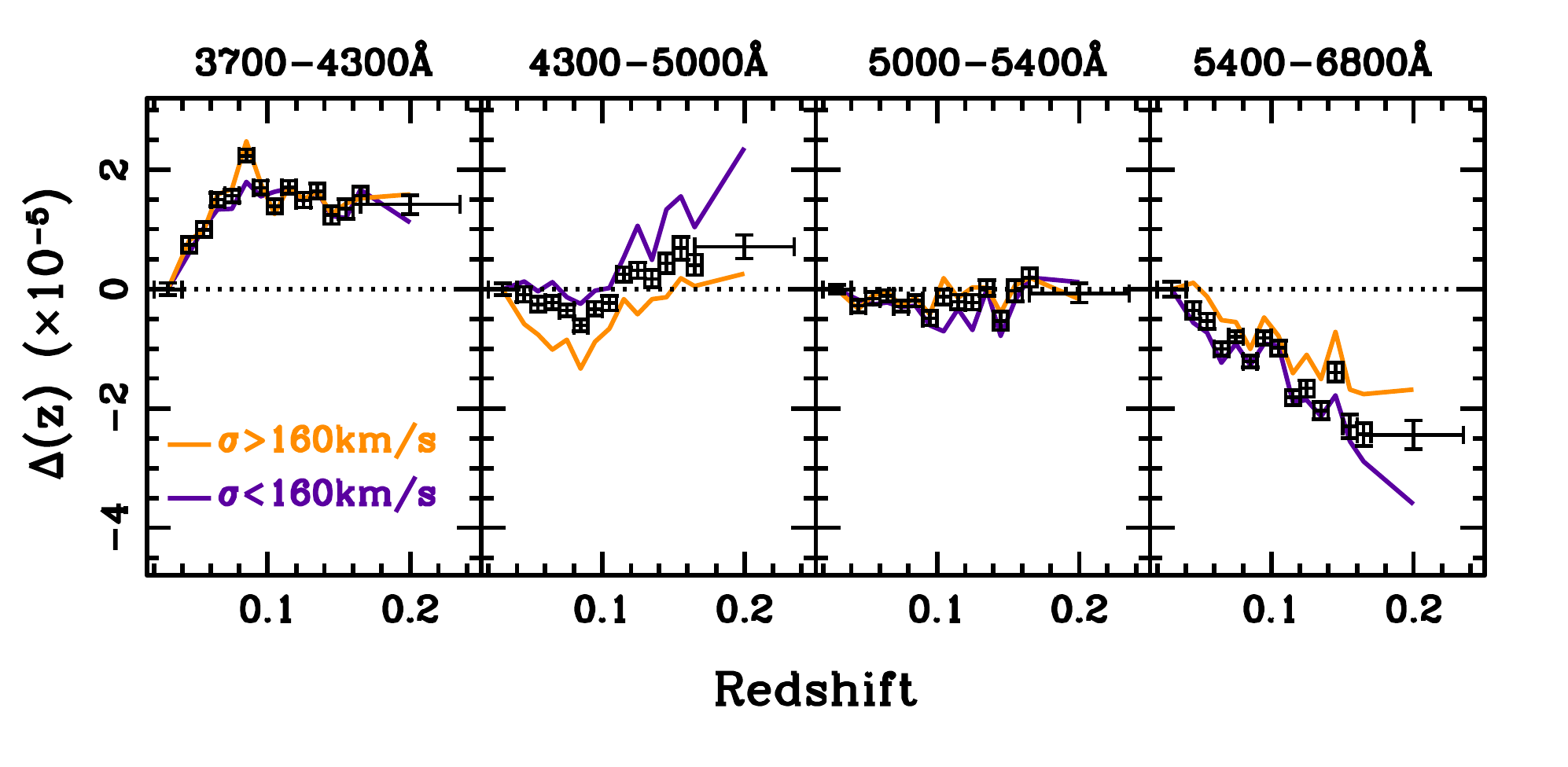}
  \caption{{\sl Top:} Illustration of the
      quantification of a possible skewness in the line fitting,
      derived from the difference in the residuals on the blue
      ($\rho_-$) and red ($\rho_+$) sides of the line.  The
      distribution of two typical lines (see Fig.~\ref{fig:Skew}) is
      shown in red and blue on the left panel. The difference in area
      between their cumulative distributions ({\sl S}, shown in
      orange, right hand panel) is used as a proxy of the
      skewness. Note the comparison is made for the same line in
      galaxies at different redshifts.  {\sl Middle:} Trend of the
      skewness parameter with redshift. The dotted lines represent the
      spectral features removed from the analysis, showed in the
      bottom panel.  {\sl Bottom:} Using only the lines with a low
      value of the skewness parameter S -- i.e. rejecting those
      corresponding to the dotted line in the panel above, we show the
      redshift and wavelength evolution of $\Delta(z,\lambda_0)$ as in
      Fig.~\ref{fig:Dz}. The trend is consistent with our previous
      findings.}
 \label{fig:Skew2}
\vskip+0.15truein
\end{figure*}

An additional systematic that one should consider is whether the shape
of the features could introduce a small effect on their derived
centroids. We emphasize that the lines being studied here are complex
superpositions of individual absorption lines blended both by the
stellar velocity dispersion of the galaxy and by the resolution of the
spectrograph. Note that for this effect to produce a signal in our
analysis, we would need a systematic trend with respect to wavelength,
as we are averaging out over several lines, comprising thousands of
spectra. Nevertheless, we test here this potential systematic. Any
departure from the Gaussian fit that we apply in our analysis will
produce an offset of the central position if the actual line is
asymmetric. Profiles such as a Lorentzian, or a Voigt profile will
yield the same central position, albeit with a different line
width. Therefore, we test a possible asymmetry (skewness) of the lines
by comparing the residuals of the fits separately on the blue and the
red side of each line. We define these residuals as follows:

\begin{align}
  \left.
  \begin{array}{llr}
    \rho_- &\equiv\frac{1}{N_-}\sum_i\left[\Phi_{\rm OBS}(\lambda_i)-\Phi_{\rm BF}(\lambda_i)\right]/\sigma_i,
    \quad & \lambda_i<\lambda_0\\
    & & \\
    \rho_+ &\equiv\frac{1}{N_+}\sum_i\left[\Phi_{\rm OBS}(\lambda_i)-\Phi_{\rm BF}(\lambda_i)\right]/\sigma_i,
    \quad & \lambda_i>\lambda_0
  \end{array}
  \right\}
  \label{eq:skew}
\end{align}

where $\Phi_{\rm OBS}(\lambda)$ and $\Phi_{\rm BF}(\lambda)$ are the
observed and best-fit spectra, respectively, and $\sigma_i$ is the RMS
uncertainty. The number of fitting points on the blue and red sides is
$N_-$ and $N_+$, respectively. A potential skewness would show up as
an asymmetry in the distribution of the residuals for a specific
line. Fig.~\ref{fig:Skew} shows a histogram of the difference,
$(\rho_--\rho_+)$ for galaxies in four redshift bins. Note that
although small variations among the distributions are apparent, there
is no significant trend that could give rise to a cumulative
$\Delta(z,\lambda)$, as the one presented in Fig.~\ref{fig:Dz}. In
order to quantify in more detail this result, we define a statistic
($S$) to measure the skewness.  \\ The cartoon in the top panels of
Fig.~\ref{fig:Skew2} illustrates the definition of $S$ for a generic
comparison: the histograms of $(\rho_--\rho_+)$ ({\sl left}) are shown
for two sets -- corresponding to the same line in samples at different
redshift. Let us assume that the red line corresponds to the
distribution at the reference redshift, whereas the blue line is the
distribution in a bin at a different redshift.  In this example, we
assume that one of the sets (the one shown in blue) has a significant
skewness.  We define our skewness-related statistic $S$ as the area
between the cumulative distributions ({\sl right}). Note that this
definition preserves the sign: the area between the cumulative
distribution of the targeted bin with respect to the reference bin is
considered positive/negative depending on whether it lies above/below
the reference. In addition, if the difference between the two sets is
symmetric about the centre (as would be the case for two Gaussians
with different widths, or a Gaussian and a Lorentzian), the cumulative
distributions will not match, but the total area, $S$ will cancel out.
This method is applied to quantify the {\sl relative} difference in
the line profiles for the same line, as a function of redshift.  Only
if this parameter correlates strongly with redshift {\sl and}
wavelength, we can expect a systematic from line shape variations to
affect our results. The middle panels of Fig.~\ref{fig:Skew2} show the
behaviour of the lines studied in this paper, binned in wavelength in
the same way as the result presented in Fig.~\ref{fig:Dz}. Note the
trend is mild, except for a few cases (dotted lines). In order to test
whether this trend could be responsible for the signal reported in
this paper, we repeated the analysis, rejecting those lines where
$|S|\gtrsim 0.02$. The bottom panel of Fig.~\ref{fig:Skew2} is the
outcome, showing that the trend found in the SDSS spectra is resilient
with respect to a skewness-related systematic.  Therefore, we conclude
that a simple measure of the asymmetry of the line profiles cannot
account for the observed variation.

\subsection{Additional Systematics}
\label{ssec:moresystematics}

\begin{figure*}
\centering
\includegraphics[width=12cm]{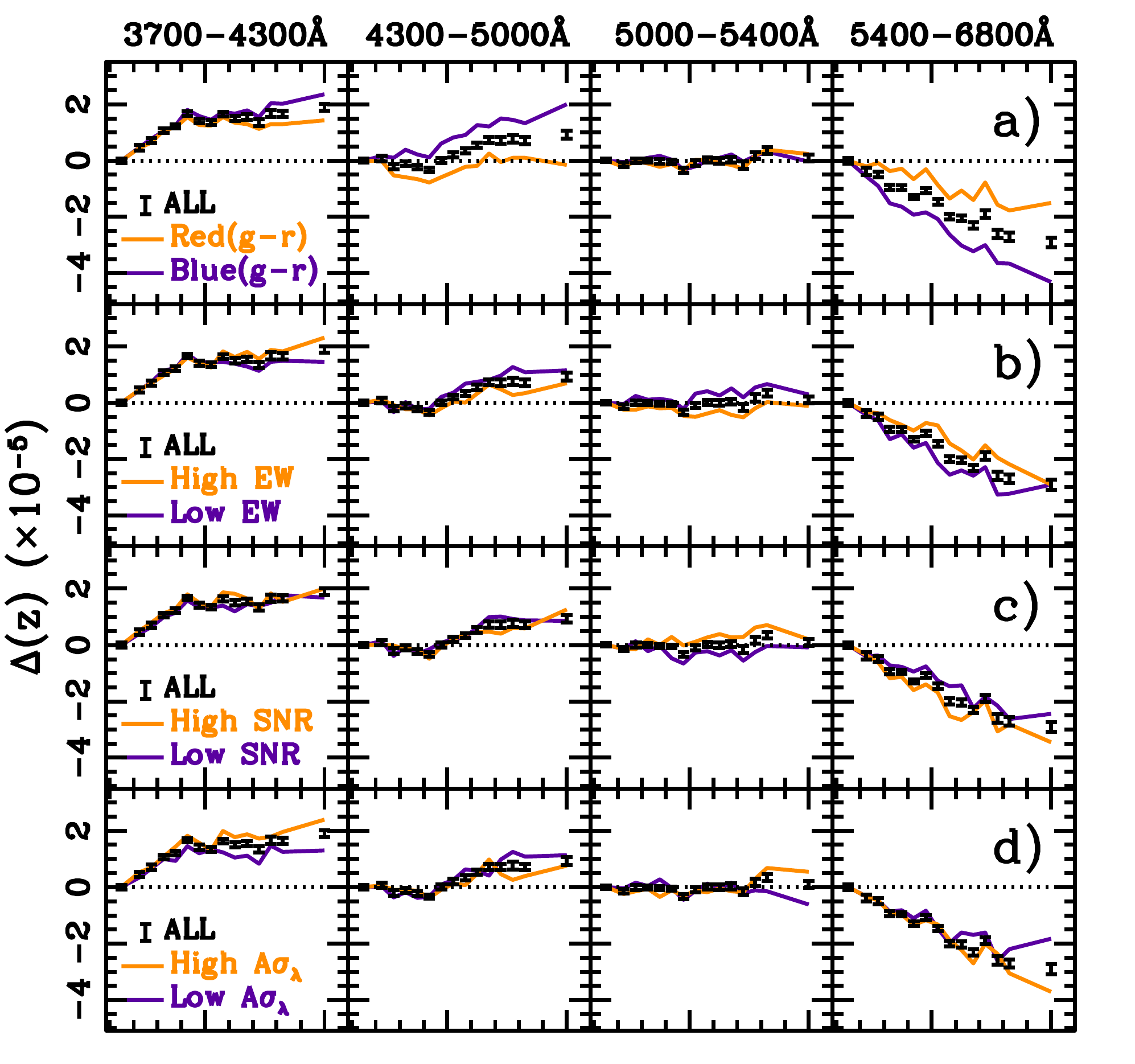}\\
\includegraphics[width=12cm]{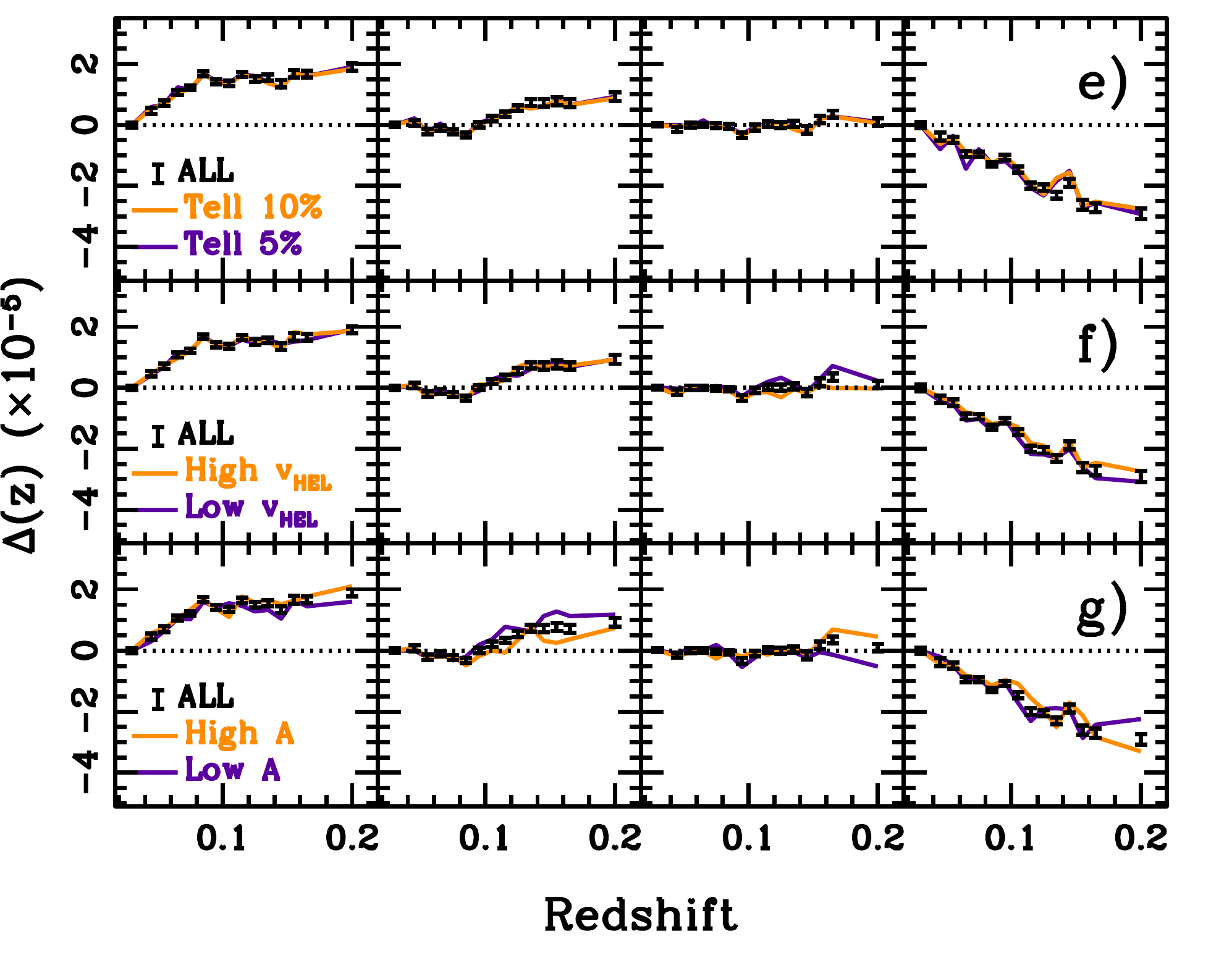}
\caption{Comparison of $\Delta(z,\lambda_0)$ for different subsamples
  of galaxy spectra. The observed departures from the standard (1+z)
  cosmological law are shown here ($\Delta(z,\lambda_0)$, as defined
  in Eq.~\ref{eq:Deltaz}). The total sample is shown as error bars
  (labelled ALL, and being identical to the sample shown in
  Fig.~\ref{fig:Dz}). The different rows show the results when
  splitting the sample at the median value of (a) (g--r) SDSS-fibre
  colour; (b) equivalent width for each of the measured spectral
  features; (c) signal-to-noise ratio; (d) total area of the Gaussian
  fit to the features; (e) after rejecting lines affected by a
  telluric correction above 5\% or 10\%, as labelled; (f) heliocentric
  velocity correction applied to the spectra; (g) Amplitude of the
  Gaussian fit. The results consistently show for all cases an offset
  from the standard (1+z) law. For ease of comparison, the vertical
  axis in all the panels extend over the same interval as in
  Fig.~\ref{fig:Dz}}
\label{fig:Dzsub}
\vskip+0.15truein
\end{figure*}

In order to further test the robustness of the signal measuring the
departure of the observed wavelengths with respect to a standard (1+z)
cosmological law, we show in Fig.~\ref{fig:Dzsub} the equivalent of
Fig.~\ref{fig:Dz} for a different set of subsamples. All panels show
the result from the complete sample -- selected in velocity dispersion
in the range 100-250\,km\,s$^{-1}$ -- as error bars, labelled
``ALL''. Row (a) shows the trend of $\Delta(z,\lambda_0)$ when
splitting the sample with respect to the SDSS-fibre colour (g--r),
after accounting for foreground Milky Way dust attenuation. For each
set of measurements within a redshift bin, we split the sample at the
median value of the distribution of (g--r).  Note this separation
creates a similar result as with respect to velocity dispersion
(Fig.~\ref{fig:Dz}) or age/metallicity of the stellar populations
(Fig.~\ref{fig:Idx} and \S\S\ref{ssec:agemet1}). However the general
trend is not removed. The next two rows follows the same procedure,
splitting the sample with respect to the equivalent width (b), and
signal-to-noise ratio (c). In (d) we consider the actual area of the
Gaussian fit ($\propto {\cal A}\sigma_\lambda$, see
eq.~\ref{eq:Fitfcn}). Row (e) explores the contribution from telluric
absorption. Here, we reject those measurements where the core of the
line -- defined as $\lambda_0\pm\sigma_\lambda$, where $\lambda_0$ is
the central wavelength of the feature and $\sigma_\lambda$ is the
width of the Gaussian fit -- is affected by a telluric correction
above 5\% or 10\%, as labelled.  To estimate the level of correction,
we use the atlas of telluric features (from Hanuschik 2006,
unpublished) and create a smoothed correction model at the spectral
resolution of the SDSS spectrograph (R$\sim$2000). Row (f) splits the
sample with respect to the heliocentric velocity correction applied to
the spectra.  Finally, we show in (g) the difference with respect to
the amplitude of the Gaussian fit (${\cal A}$ in eq.~\ref{eq:Fitfcn}).
Note that in all cases the trend presented in Fig.~\ref{fig:Dz} is
reproduced.

\begin{figure*}
\centering
\includegraphics[width=14cm]{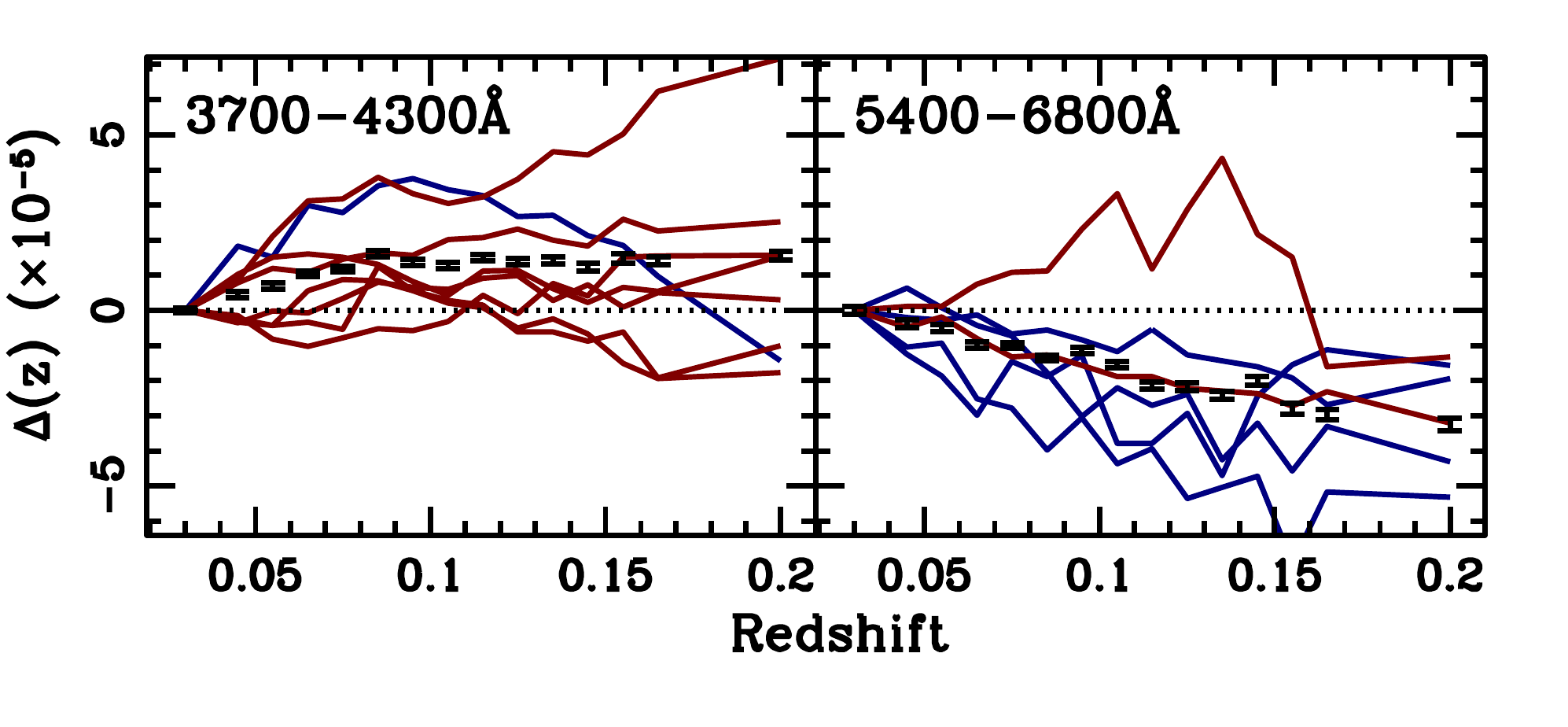}\\
\includegraphics[width=14cm]{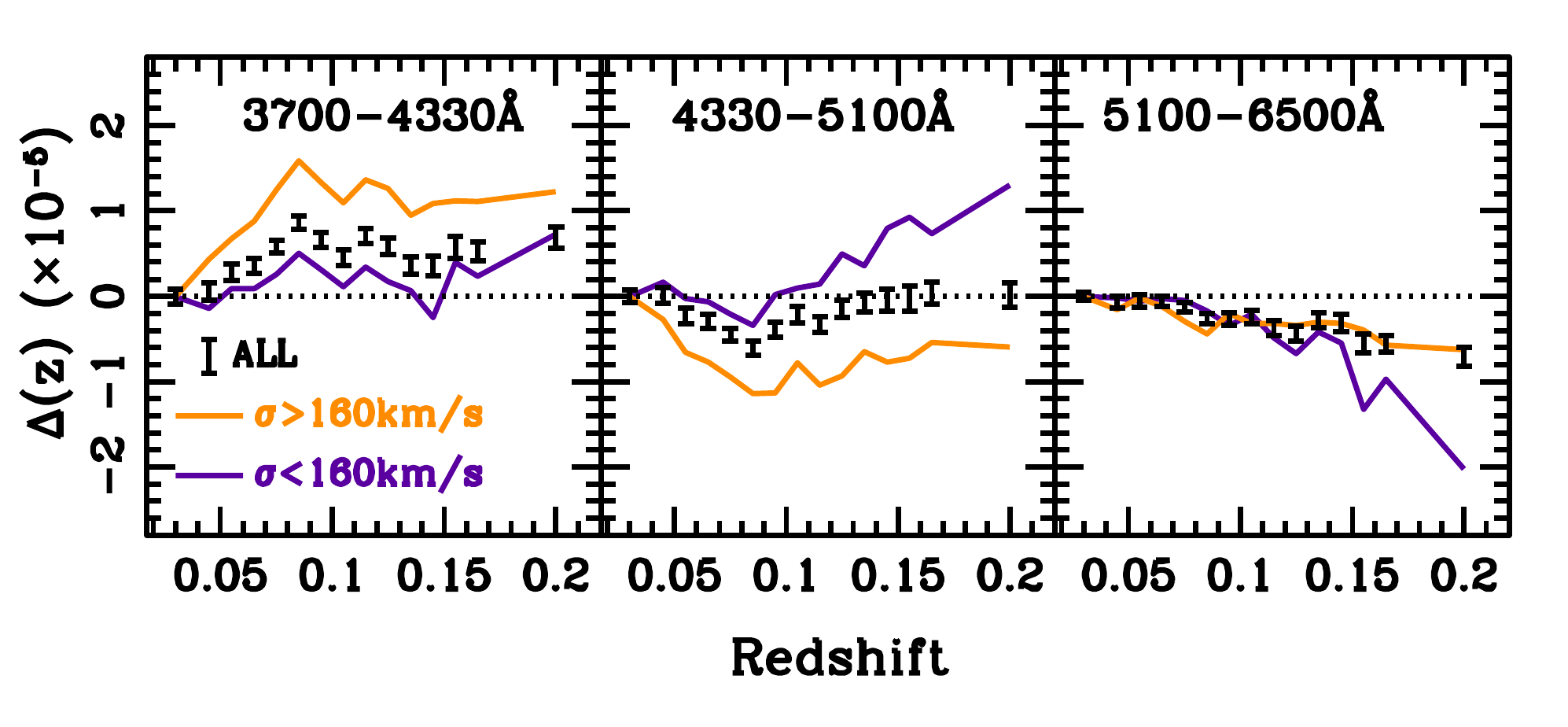}
\caption{Top: Departure from the standard (1+z) law for individual
  spectral lines. We show here the trend of $\Delta(z;\lambda_0)$ vs redshift
  for the complete sample (error bars), along with the result for the
  fits for individual spectral features within the rest-frame spectral
  window given in each panel. Each line is colour coded depending on
  whether the feature is mostly in emission ({\sl blue}) or absorption
  ({\sl red}). Bottom: Trend of $\Delta(z)$ with redshift, when all
  the emission lines are removed from the analysis. Note we now split
  the reduced spectral range in three parts.  }
\label{fig:DzLines}
\vskip+0.15truein
\end{figure*}

Fig.~\ref{fig:DzLines} ({\sl top}) shows a similar plot, where each
line represents the trend estimated for a single spectral feature. The
lines are colour-coded, with blue lines representing features mostly
in emission, and red lines corresponding to absorption features. For
reference, the error bars show the results presented in
Fig.~\ref{fig:Dz}, which represent the median extended to all
measurements from all galaxies and all lines within the rest-frame
spectral window given in each panel. Note that aside from the large
scatter expected from individual lines, there is a clear trend between
the blue- ({\sl left}) and the red-end ({\sl right}) of the spectral
window.  We point out that the information in the blue end is
dominated by absorption lines (except for [OII]), whereas at the red
end, the situation is reversed, and the emission lines from H$\alpha$,
[NII], [SII] contribute the most to this wavelength bin. In order to
assess whether the observed departure from the (1+z) law is related to
the difference between absorption and emission (i.e. tracing the
difference between the gas and the stellar component of the galaxies),
we show in the bottom panel of Fig.~\ref{fig:DzLines} the same
analysis as in Fig.~\ref{fig:Dz}, restricted only to absorption
lines. Although in this occasion we need to split the spectral range
in three bins, given the reduced set of lines, the result is
nevertheless robust against this issue.

\begin{figure*}
\centering
\includegraphics[width=14cm]{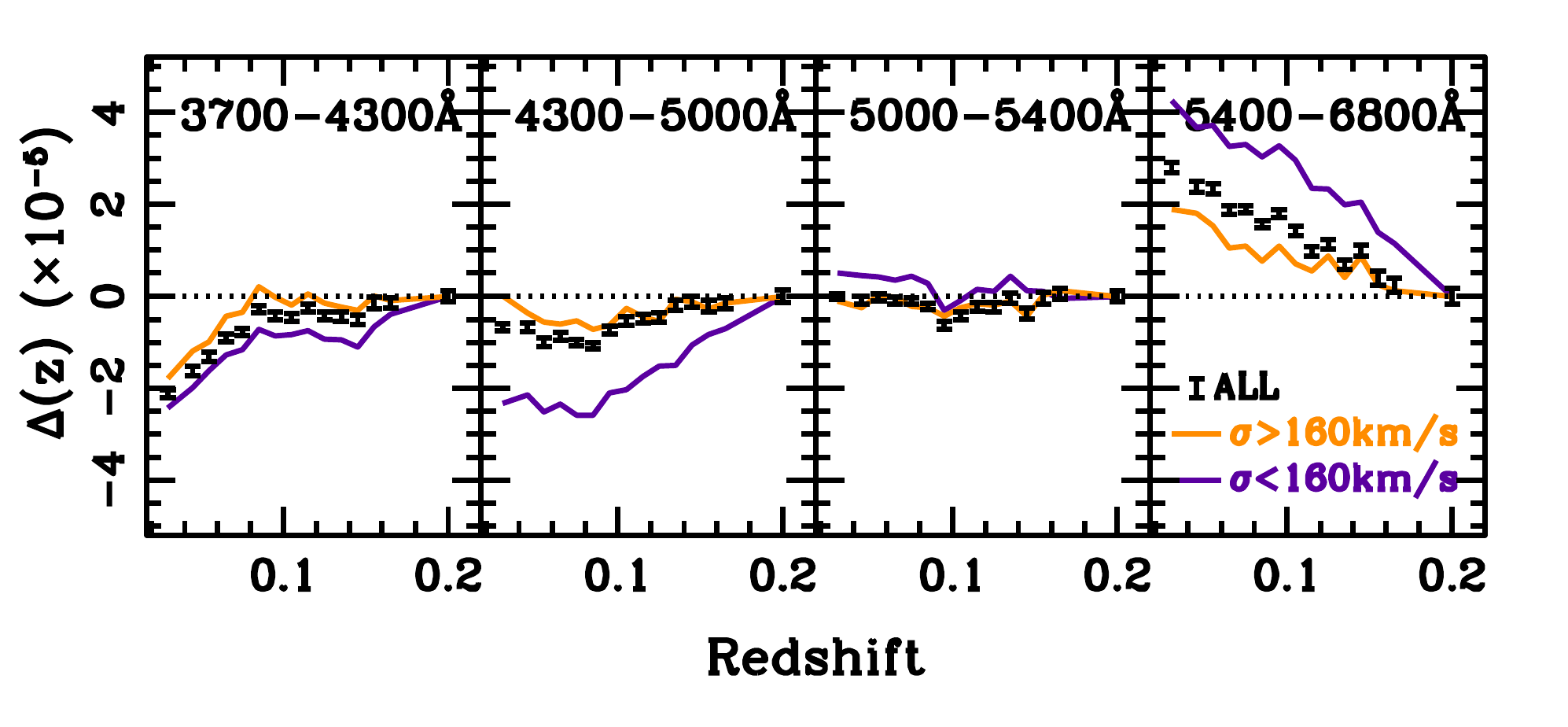}
\caption{Trend of $\Delta(z)$ with redshift as in Fig.~\ref{fig:Dz},
with the reference redshift changed to the highest redshift bin
(z=0.18-0.25).}
\label{fig:DzLines_z0200}
\vskip+0.15truein
\end{figure*}

Finally, Fig.~\ref{fig:DzLines_z0200} shows the same analysis, when
changing the reference redshift -- which is needed to define the
position of the spectral features -- from z=0.02-0.04 (used throughout
this work) to the highest redshifts, z=0.18-0.25.  Such a change
does not affect the result either, except for the expected change in
the sign of $\Delta(z)$.

\bibliography{SDSSz_FT}

\end{document}